\documentclass[conference]{IEEEtran}
\IEEEoverridecommandlockouts

\usepackage[style=ieee,backend=biber]{biblatex}
\usepackage{caption}
\usepackage{graphicx}
\usepackage{textcomp}
\usepackage{hyperref}
\usepackage{fancyref}
\usepackage{placeins}
\usepackage{xcolor}
\usepackage{float}
\usepackage{amsmath}

\title{Investigating the Suitability of Delay Tolerant Networks for Broadcasting Tsunami Warnings in Palu, Indonesia}

\author{\IEEEauthorblockN{Adam Graham}
\IEEEauthorblockA{\textit{School of Computer Science} \\
\textit{The University of Nottingham} \\
psyag11@nottingham.ac.uk}
\and
\IEEEauthorblockN{Milena Radenkovic}
\IEEEauthorblockA{\textit{School of Computer Science} \\
\textit{The University of Nottingham} \\
milena.radenkovic@nottingham.ac.uk}}

\begin{document}

\maketitle

\begin{abstract}
On the 28th of September, 2018, a tsunami hit the city of Palu in Indonesia, killing 4,340 people. The earthquake preceding the tsunami crippled communication lines and may have rendered the transmission of tsunami warning messages using traditional end-to-end approaches impossible. This paper proposes an alternative approach using Delay Tolerant Networks (DTNs) for tsunami warning message routing given their resilience to disruptions and sparse connections. Both Epidemic and Spray and Wait routing protocols were simulated in a pseudo-realistic environment to evaluate their effectiveness for transmitting tsunami warning messages in Palu. Results indicated that these protocols are not suitable for the tight time constraints of post-earthquake tsunami warnings with the currently available technology. However, they may have promising applications for the earthquakes that precede tsunamis.\\
\end{abstract}


\section{Mobile ad hoc Delay Tolerant Networks and the ONE Simulator} \label{section-1}

Mobile Ad hoc Networks (MANETs) are wireless, infrastructure-less networks that are formed solely by independent, self-organising devices \parencite{basagni2004mobile}. Nodes within MANETs can communicate with other nodes that are directly within their transmission range, however communication with nodes outside of their transmission range requires the use of a multi-hop routing approach \parencite{hoebeke2004overview}. This approach entails intermediate nodes acting as routers to discover routes of nodes that the message can traverse by hopping between nodes \parencite{basagni2004mobile} in order to reach its destination. As MANETs adopt TCP/IP for end-to-end communication \parencite{ABOLHASAN20041} an end-to-end route is therefore required to facilitate communication \parencite{Vasilakos2025-zk}. This requirement becomes problematic since the nodes within MANETs move arbitrarily and unpredictably, resulting in frequent route changes and network partitions \parencite{CHLAMTAC200313}. Consequently, MANETs can struggle to handle unforeseen delays and disruptions in communication, which could become detrimental if they are used in operations like emergency services. Despite this, MANETs still offer a low cost solution to networking while being easy to deploy \parencite{basagni2004mobile}. Additionally, their lack of infrastructure allows them to be a feasible option for remote geographical locations where traditional networking may be impractical due to the necessary infrastructure being unavailable.\\

\begin{figure}[H]
    \centering
    \includegraphics[width=0.52\linewidth]{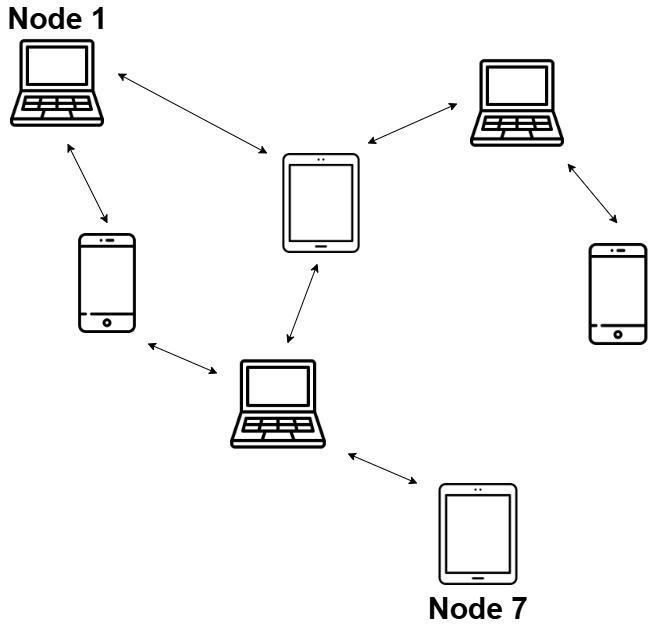}
    \caption{A visual representation of a MANET, produced by the author with credit to \parencite{iconixar2025} for the node icons. Note that a message sent from node 1 would take three hops to reach node 7.}
    \label{fig:manet}
\end{figure}

Delay Tolerant Networks (DTNs) are able to handle disruptions during transmission and long delays \parencite{Vasilakos2025-zk} through the use of a store-carry-forward approach \parencite{8226101}. Unlike synchronous end-to-end transmission, this approach involves asynchronous message transmission in which the message hops across nodes with the nodes hopped to also storing the message \parencite{8226101}. Since intermediate nodes have stored the message, they are able to wait as long as they need before forwarding the message to the next node, which becomes necessary in intermittent environments \parencite{e100-b_1_2}. This allows DTNs to outperform MANETs in sparse networks \parencite{6402866} and excel in environments where connections are inconsistent. They could have promising applications in aerospace, where node contact is infrequent, and natural disaster situations, where traditional communication may become impaired. However, since DTNs are designed to eventually deliver messages \parencite{10.1145/1015467.1015484} they may become disadvantageous in situations where faster network approaches are available. As DTNs do not transmit messages based on an end-to-end path, the end-to-end protocols that are traditionally used in networks will not work and new protocols are required \parencite{Vasilakos2025-zk}. Some examples of protocols used with DTNs include: First Contact, Epidemic, Spray and Wait, PRoPHET, and MaxProp \parencite{BENAMAR2014141}.\\

The Opportunistic Network Environment (ONE) simulator is a Java-based event simulation engine, designed to simulate and evaluate DTN routing \parencite{keranen-theone}. Users can configure ONE to simulate their own scenarios with text-based settings files that the simulator uses to determine which Java classes it should use and the values it should assign to their fields \parencite{keranen2008opportunistic}. For example, if a user wants every node to be routed using a spray and wait router with an upper limit of six message copies, then they would include \textit{Group.router = SprayAndWaitRouter} and \textit{SprayAndWaitRouter.nrofCopies = 6}. This tells ONE to use the \texttt{SprayAndWaitRouter} Java class and set its \texttt{nrofCopies} field to 6. By default, ONE provides six routing protocols: Direct Delivery, First Contact, Spray and Wait, PRoPHET, MaxProp, and Epidemic; however, users can implement additional protocols \parencite{keranen-theone}. One of ONE's greatest strengths is the level of abstraction it provides, as users are not required to understand or modify ONE's codebase in order to carry out comprehensive simulations and experiments for their research. This makes the simulator highly accessible since users are not expected to have prerequisite programming knowledge. Additionally, ONE is able to run simulations on both real-world and custom-made maps \parencite{keranen-theone}. This enables users to simulate and experiment with pseudo-realistic data, which would help produce more representative results. Despite this, the ONE simulator runs simulations using a fixed time slice approach, meaning that time is advanced in time steps rather than being continuous, like in the real world \parencite{keranen2008opportunistic}. As a result, the simulator only recognises events happening at either the start or end of a time step \parencite{keranen2008opportunistic}. Consequently, any experiments carried out in the ONE simulator will forfeit precision and realism, proportionate to how long the duration of each time step is. Time steps can be made smaller to combat this, but doing so will make the simulation run much slower \parencite{keranen2008opportunistic} so the extent to which this would solve the issue is limited.\\

\section{The scenario} \label{section-2}

Since 1907, Indonesia has suffered 368,904 deaths and \$14,894.75 million in damages from tsunami's \parencite{ncei_tsunami_database}. A notable example of this is the tsunami that hit Palu City on the 28th of September, 2018 \parencite{SABAH2023474} which resulted in 4,340 deaths and \$1,500 million in damages \parencite{ncei_tsunami_database}. A magnitude 7.5 earthquake caused this tsunami \parencite{SABAH2023474} the epicentre was located 70km north of Palu City and radar readings showed a 170km long rupture that passed through the west side of Palu City \parencite{palu_tsunami_insights}. The earthquake brought down Palu's power and communication lines which may have prevented the tsunami alerts from reaching residents \parencite{BBC2018}. This paper proposes a DTN networking approach to transmitting a warning message from a seismic station to a local disaster management agency (BPBD \parencite{amrullah2023challenges}) office. The aim of this approach is to provide a solution that can operate as a reliable method for relaying tsunami alerts, even while other forms of local communication are impaired.\\

\begin{figure}[H]
   \centering
   \includegraphics[width=0.9\linewidth]{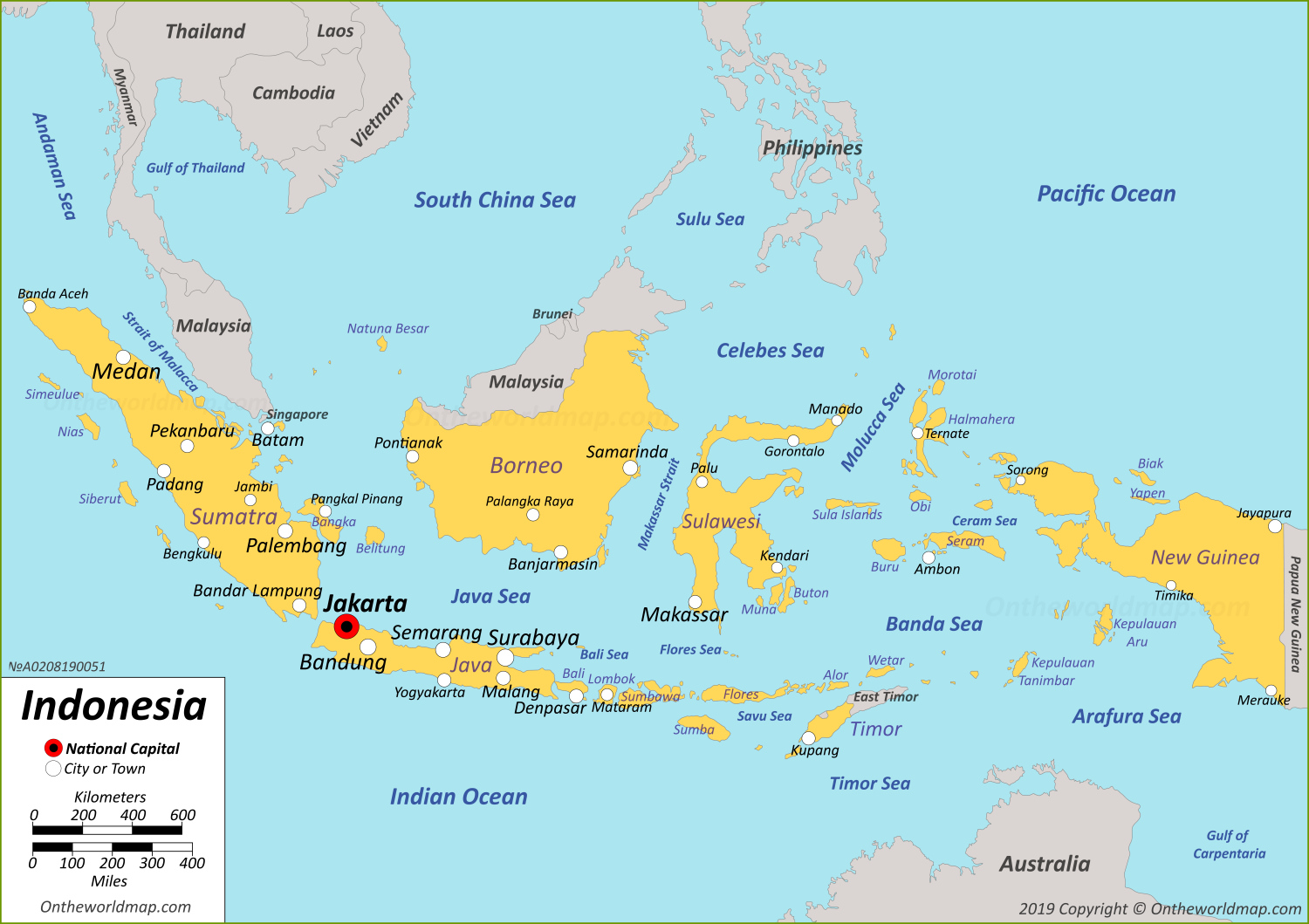}
   \caption{A map of Indonesia, highlighted in orange, with Palu located in Sulawesi \parencite{Ontheworldmapcom2019}.}
   \label{fig:indonesia-map}
\end{figure}

\begin{figure}[H]
   \centering
   \includegraphics[width=0.745\linewidth]{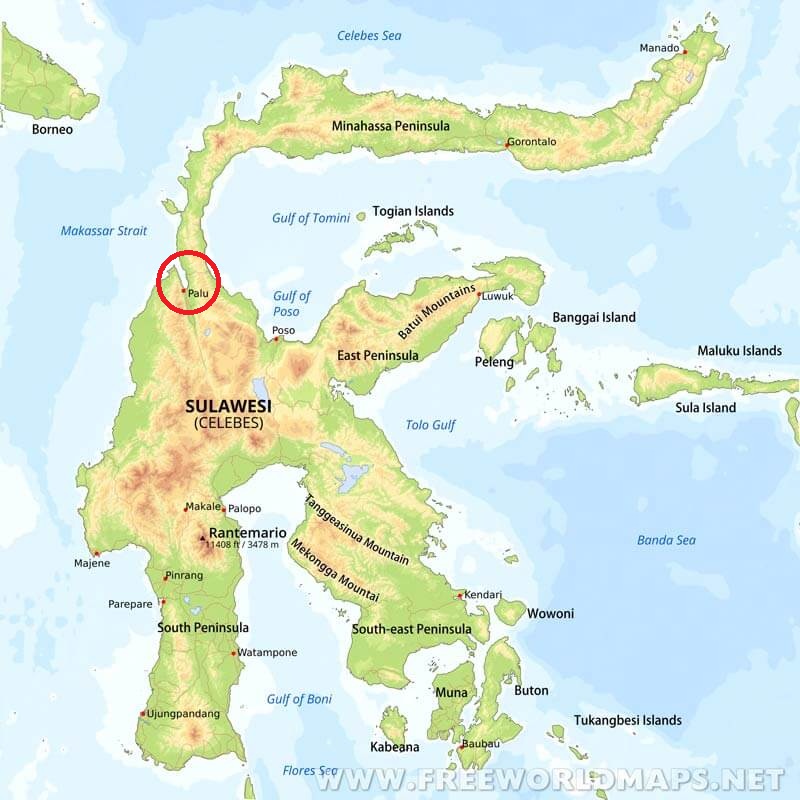}
   \caption{A map of Sulawesi with Palu circled, adapted from \parencite{Freeworldmaps}.}
   \label{fig:sulawesi-map}
\end{figure}

\begin{figure}[H]
   \centering
   \includegraphics[width=0.685\linewidth]{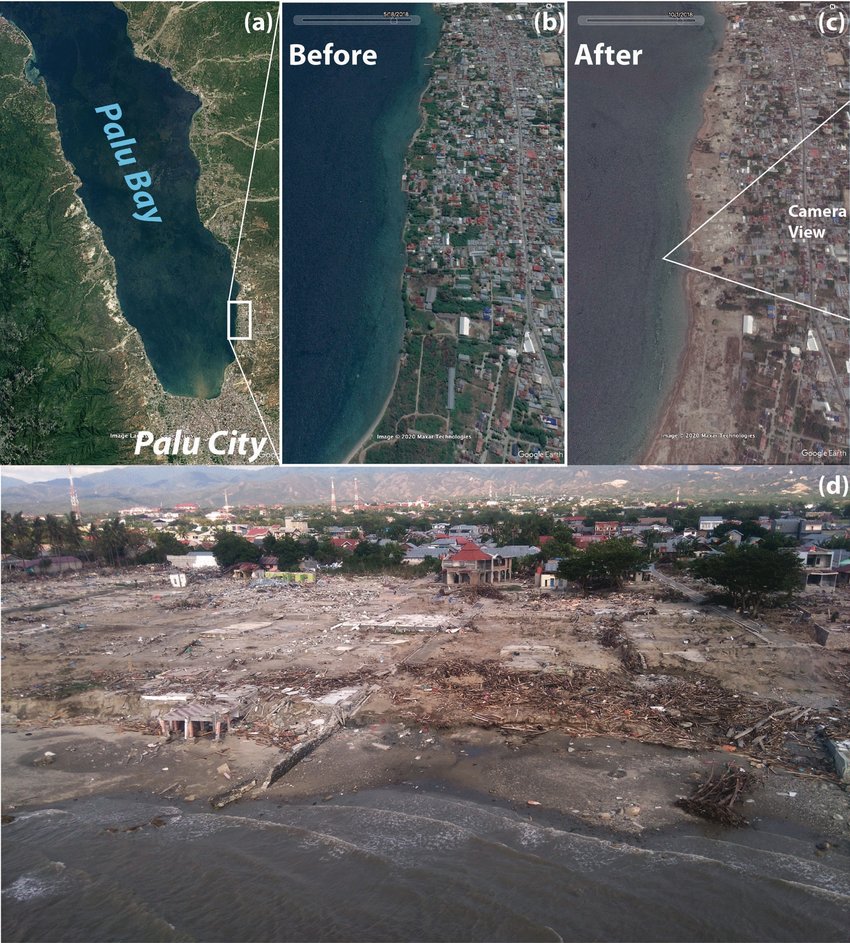}
   \caption{A visual comparison of Palu before and after the tsunami \parencite{palu_before_after}.}
   \label{fig:tsunami-impact}
\end{figure}

\section{Investigation on the design of Epidemic and Spray \& Wait} \label{section-3}

Epidemic \parencite{vahdat2000epidemic} is a replication-based routing protocol \parencite{Burleigh2020} that involves both nodes sharing all the messages they do not have in common when two nodes meet \parencite{gamit2014evaluation}. This results in the nodes repeatedly replicating and sharing the message that the source is attempting to transmit to the recipient \parencite{alaoui2015performance}, causing the message to quickly spread across connected parts of the network \parencite{vahdat2000epidemic}.\\

\begin{figure}[h!]
    \centering
    \includegraphics[width=0.721\linewidth]{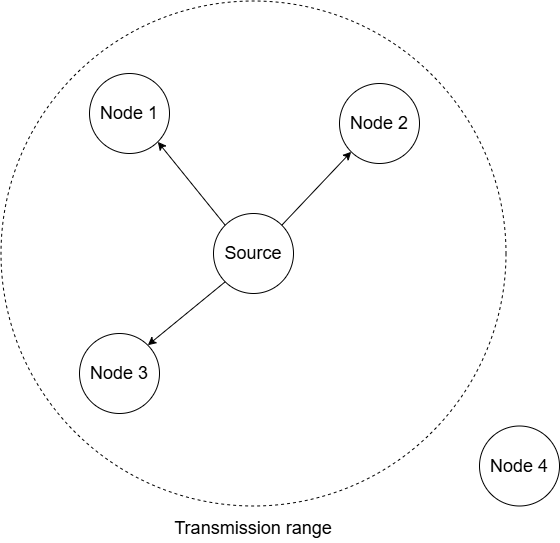}
    \caption{A visual representation of epidemic, produced by the author.}
    \label{fig:epidemic}
\end{figure}

Spray and wait \parencite{spray_and_wait_original} is a flooding (replication) based routing protocol \parencite{spray_and_wait_original} which limits the maximum number of copies that a message can have, where the maximum number of copies is denoted by the parameter \textit{L} \parencite{alaoui2015performance}. Spray and wait consists of two phases, a spray phase and a wait phase \parencite{spray_and_wait_original}. During the spray phase, every message at the source is forwarded such that each message has a copy stored in \textit{L} distinct nodes \parencite{alaoui2015performance}. Each of the \textit{L} nodes, including the source, holds a single copy of all the source's messages \parencite{Burleigh2020}, assuming they have sufficient buffer space. Note that the nodes which receive the source's messages are the first \textit{L-1} nodes that the source encounters \parencite{Burleigh2020}. After the spray phase has finished the wait phase begins. If the destination node did not receive the message being transmitted then each of the \textit{L} nodes focus on direct transmission \parencite{spray_and_wait_original} and wait until they encounter the destination node to directly transmit the message to it \parencite{gamit2014evaluation}.\\\\\\\\\\\\\\\\\\\\

\begin{figure}[h!]
    \centering
    \includegraphics[width=0.721\linewidth]{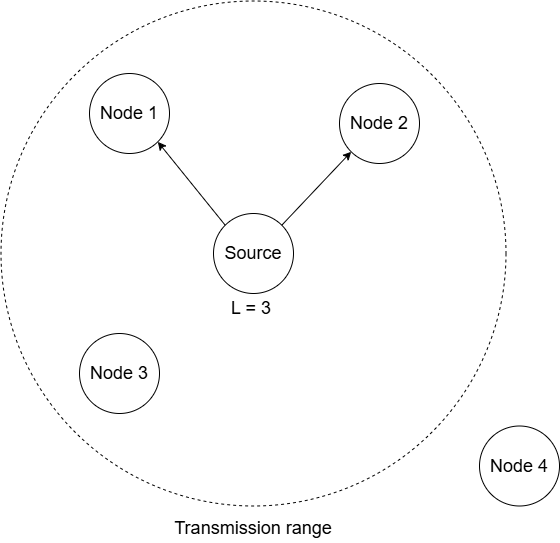}
    \caption{A visual representation of the spray phase in spray and wait, produced by the author.\\}
    \label{fig:spray-phase}
\end{figure}

\begin{figure}[h!]
    \centering
    \includegraphics[width=0.721\linewidth]{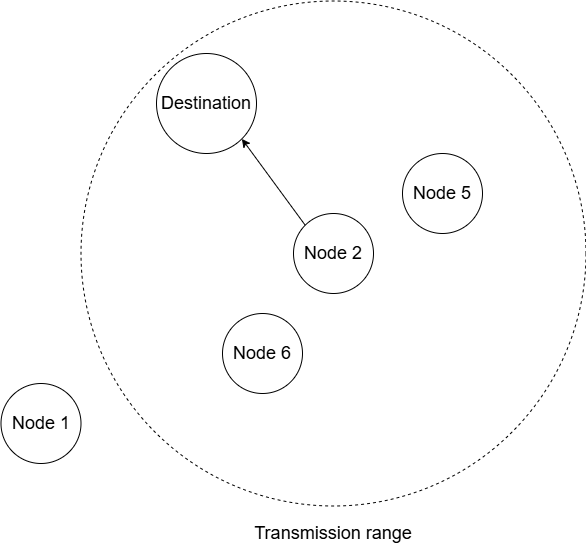}
    \caption{A visual representation of the wait phase in spray and wait, produced by the author.}
    \label{fig:wait-phase}
\end{figure}

Epidemic is suitable for this paper's scenario as it minimises latency and maximises delivery rate \parencite{vahdat2000epidemic}. This is important as based on \parencite{geist2006tsunami, gonzalez1999tsunami, dudley1998tsunami} tsunamis can take anywhere from a few minutes to a couple of hours to arrive, and the first wave of the tsunami that hit Palu only took 30 minutes to arrive \parencite{most_destructive_tsunamis}. Tsunamis are time-sensitive so delayed and/or unsuccessful delivery attempts could result in people not receiving warning messages in time, and potentially catastrophic consequences. However, epidemic is prone to consuming high amounts of node buffer space and network bandwidth \parencite{Burleigh2020}. This risks filling up node buffers too much during transmission, which could prevent nodes from picking up subsequent messages from the BPBD office, like warning messages and evacuation instructions.\\

Spray and wait is also suitable for this scenario since it aims to reduce the number of transmissions required for a message to reach its destination and minimise the delivery delay \parencite{spray_and_wait_original}. Therefore, spray and wait reduces the risk of node buffers filling up too much during transmission which ensures they are ready to receive and transmit subsequent messages. It also achieves good scalability, which enables it to perform well in varying levels of connectivity and network size \parencite{spray_and_wait_original}. This could allow the proposed approach to scale beyond use in Palu, potentially to larger geographical areas that face similar problems with tsunamis. Despite these, the direct transmission of the wait phase does rely on at least one of the \textit{L} nodes encountering the destination in order for the message to be delivered \parencite{kim2010composite}. For a tsunami situation with tight time restrictions, there is a risk that none of the \textit{L} nodes encounter the destination in time, especially if there's a large distance between the source and destination.\\

Implementations for both epidemic and spray and wait are already included in the ONE simulator \parencite{keranen-theone}, with specific details on their implementations being available in the ONE code repository \parencite{github_repo} and the Javadocs \parencite{javadocs}. The following implementation descriptions are based on the code in the \texttt{EpidemicRouter}, \texttt{SprayAndWaitRouter}, and \texttt{ActiveRouter} Java classes in the code repository. Both \texttt{EpidemicRouter} and \texttt{SprayAndWaitRouter} inherit from the \texttt{ActiveRouter} class. Every node in the ONE simulator has its own router that determines which protocol the node will follow when transferring messages. At each time step, \texttt{EpidemicRouter}: checks if a message transfer is already in progress, checks whether any messages can be sent to the destination, and then attempts to send all messages to every node that it's connected to (encounters). \texttt{SprayAndWaitRouter} works similarly to \texttt{Epidemic router} in how it checks whether a message is already being transferred or if a message can be sent to a destination first. However, \texttt{SprayAndWaitRouter} then creates a list of all its messages that still have at least one copy that can be shared to other nodes; simply put, any messages that have been forwarded less than \textit{L-1} times. It then attempts to transmit all of the messages in this list to every node it's connected to.\\

\section{The Design and Set-up of Experiments} \label{section-4}

The simulator was set up with a pseudo-realistic map that reflected the real-life layout of Palu city, exported from OpenStreetMap \parencite{OpenStreetMap}. This map was used for the simulations of all the following experiments.\\

\begin{figure}[h!]
    \centering
    \includegraphics[width=1.05\linewidth]{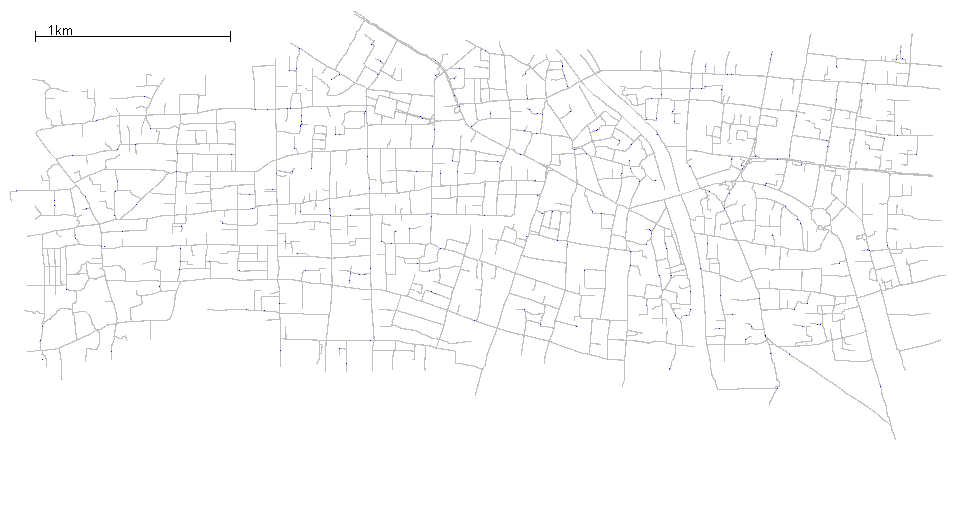}
    \caption{A screenshot of how the map looks in the ONE simulator.}
    \label{fig:palu-in-one}
\end{figure}

As discussed in section \ref{section-3}, the tsunami that hit Palu only took 30 minutes to arrive \parencite{most_destructive_tsunamis} and many other tsunami's arrived in less than an hour. In an attempt to reflect realistic time scales for a tsunami situation, the simulation was set up to run for 30 minutes of simulation time. Three groups of nodes were identified: a seismic station, a BPBD office, and cars travelling around Palu city. The seismic station and BPBD building were static nodes representing the source and destination respectively. Their positions in the simulated map reflect their geographical locations in the real world, with the real world position of the seismic station being located with the BMKG seedlink monitor \parencite{BMKG2022}, and the real-world position of the BPBD office being located using the location from the Google Maps API \parencite{GoogleMaps2025} on the official website for the BPBD in Palu \parencite{BPBD2025}.\\

\begin{figure}[H]
    \centering
    \includegraphics[width=1.0\linewidth]{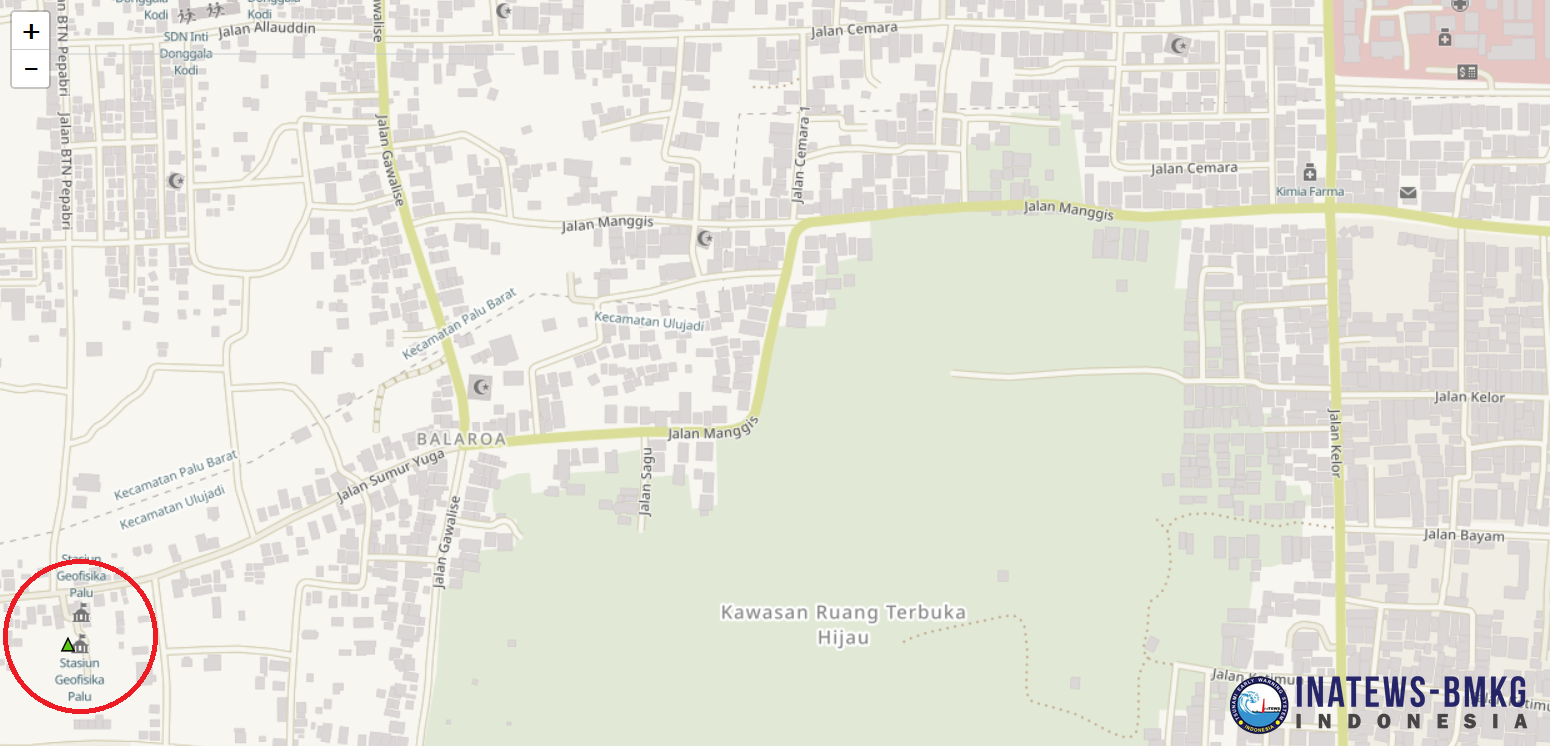}
    \caption{The seismic station's actual geographical location.}
    \label{fig:seismic-actual}
\end{figure}

\begin{figure}[H]
    \centering
    \includegraphics[width=1.0\linewidth]{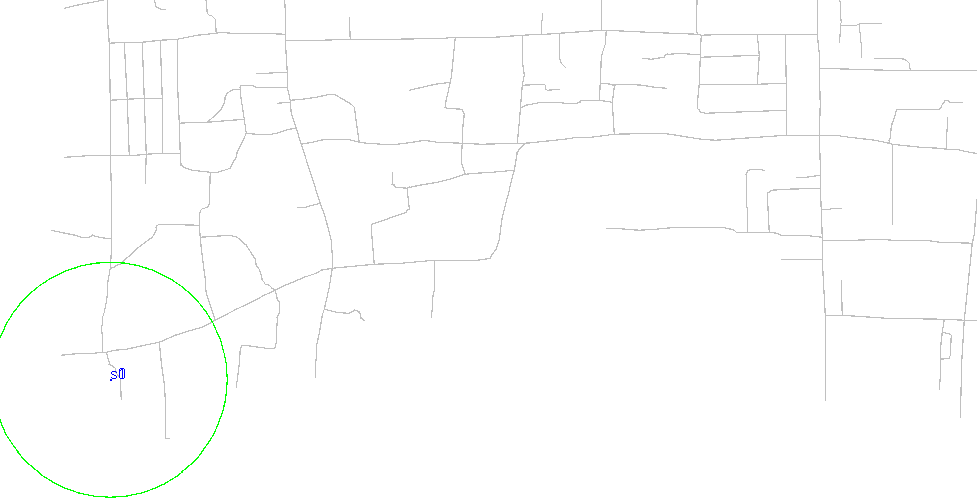}
    \caption{The seismic station's location in the simulated map.}
    \label{fig:seismic-simulated}
\end{figure}

\begin{figure}[h!]
    \centering
    \includegraphics[width=1.0\linewidth]{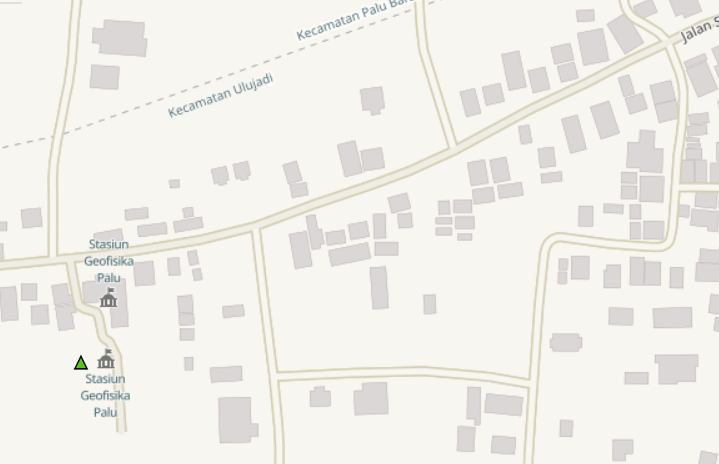}
    \caption{A magnified view of the seismic station's actual geographical location.}
    \label{fig:magnified-seismic-actual}
\end{figure}

\begin{figure}[h!]
    \centering
    \includegraphics[width=1.0\linewidth]{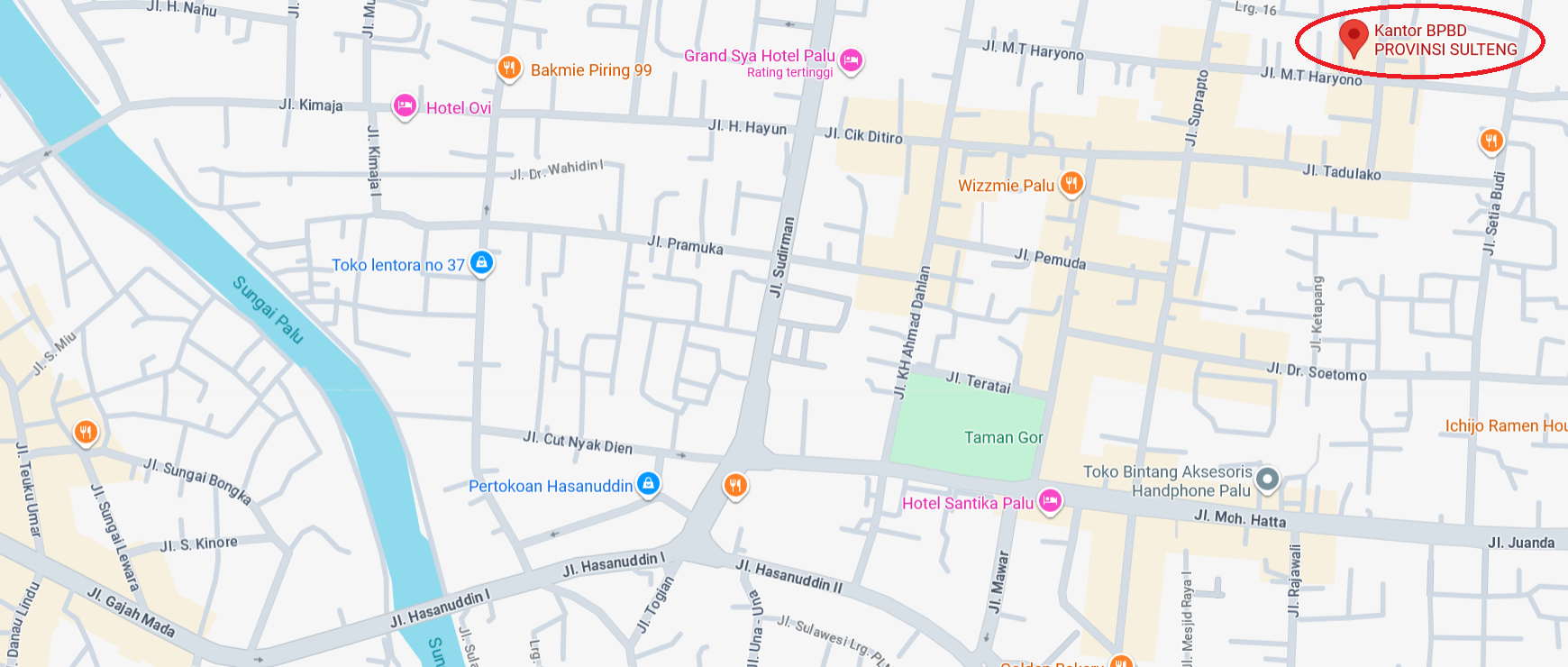}
    \caption{The BPBD office's actual geographical location.}
    \label{fig:bpbd-actual}
\end{figure}

\begin{figure}[H]
    \centering
    \includegraphics[width=1.0\linewidth]{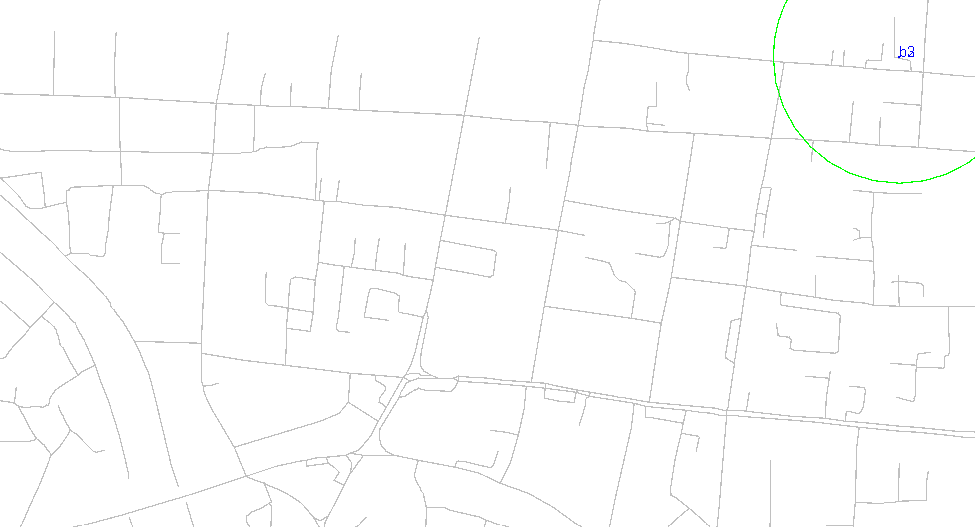}
    \caption{The BPBD office's location in the simulated map.}
    \label{fig:bpbd-simulated}
\end{figure}

\begin{figure}[h!]
    \centering
    \includegraphics[width=1.0\linewidth]{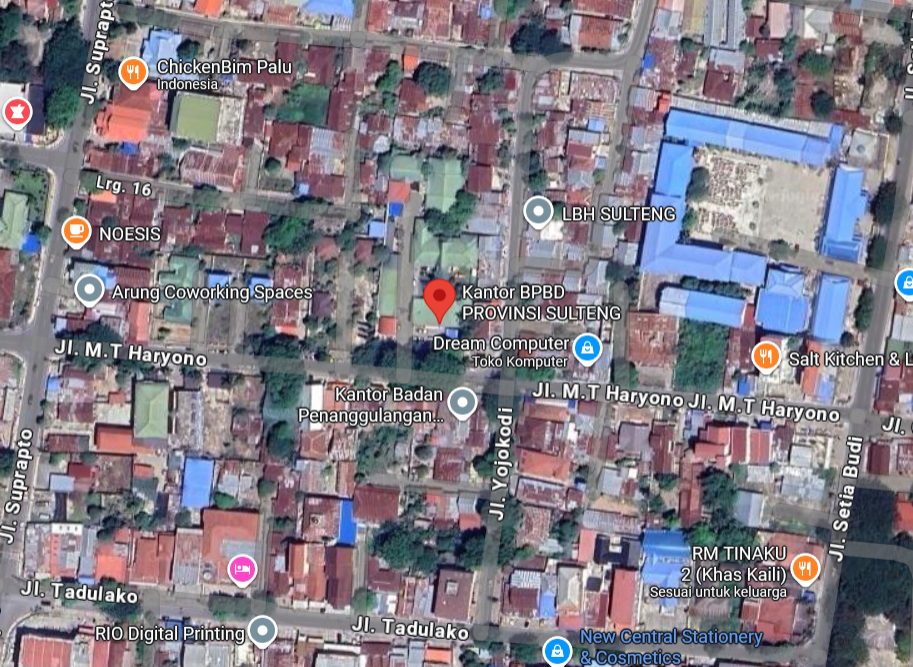}
    \caption{A magnified, satellite view of the BPBD office's actual geographical location.}
    \label{fig:bpbd-satellite}
\end{figure}

The number of car nodes used in the simulation was calculated using a variety of statistics. In 2022, Indonesia was estimated to have 17,168,862 passenger cars \parencite{IndonesiaStatePolice2022}, an estimated population of 275,773,800 by the middle of the year \parencite{InterimPopulationProjection2022}, and Palu had a population of 381,572 \parencite{PaluCityCentralStatisticsAgency2023}.

\begin{gather*} \label{eq1}
(381,572 / 275,773,800) * 17,168,862 \\
= 23755.5453457 \approx 23,756
\end{gather*}

Using these statistics and the equation above, it can be reasonably approximated that there were 23,756 cars in Palu in 2022. Given the computational limits of the ONE simulator, experiments were modelled with each node representing 100 cars.\\

In order to simulate a variety of environments, the experiments were repeated across multiple iterations, each with different parameter settings. The first set of experiments focused on the parameters of the transmission interface that facilitated communication between nodes; this paper's simulations used Bluetooth as the transmission interface. Two versions of Bluetooth were considered for these experiments, version 4.0 and version 5.0 since they were the most recent full versions, released in 2010 and 2016 respectively \parencite{al2023bluetooth}. It is worth mentioning that Bluetooth 6.0 was released in 2024 \parencite{koulouras2025evolution}, but at the time of these experiments, it was deemed unreasonable to expect widespread usage and infrastructure little over a year after its release. The experiments considered transmission speed and transmission range since these factors influence the speed and frequency of communication. Message size was trialled in later experiments that focused on protocol efficiency.  Bluetooth version 4 had a 50 to 100 metre outdoor range and a maximum transmission speed of 1 Mbps \parencite{al2023bluetooth}. Meanwhile, Bluetooth version 5 had a 200 metre outdoor range and a maximum transmission speed of 2 Mbps \parencite{al2023bluetooth}. Therefore, the transmission interface experiments were split across two iterations, the first measured the performance of both routing protocols with respect to Bluetooth version 4. The second iteration measured the performance of both routing protocols when Bluetooth version 5 was used. To model Bluetooth version 4, the first iteration used a transmission speed of 1 Mbps, which the ONE simulator measures as 125kBps, and a transmission range of 100. The second iteration modelled Bluetooth version 5 with a transmission speed of 2 Mbps (250kBps) and a transmission range of 200. Later experiments, that did not test the transmission interface, used Bluetooth version 5's parameter settings by default.\\

The next set of experiments focused on testing how well the routing protocols handled varying intensities of network resource restrictions. As mentioned in section \ref{section-3}, receiving subsequent messages is critical in an emergency situation with minimal time available. Therefore, protocol efficiency is an important metric to trial. Message size and car node buffer size were changed throughout the experiments across multiple iterations, similarly to how transmission speed and transmission range were tested. Recently in Indonesia, interpretations of seismic readings for web viewing were represented in HVSR (H/V Spectral Ratio) and PSD (Power Spectral Density) graphics with resolutions of 900x900 pixels and 640x480 pixels respectively \parencite{yuliatmoko2023seismic}. While these were designed for web viewing, for this paper's experiments they were deemed reasonable for modelling contents of a warning message in a pseudo-realistic simulated scenario since they could aid the BPBD in deciding on a course of action. By using a calculator designed to calculate image file size \parencite{KennethAlambra2024}, it was calculated that - with a 16-bit bit depth - the HVSR graphic would be approximately 1.62Mb and the PSD graphic would be approximately 0.6144Mb. While these calculations assumed that the images were not compressed, which is unlikely to be the case for a message being transferred wirelessly, they were exported as PNGs (Portable Network Graphics \parencite{pngs}) \parencite{yuliatmoko2023seismic} which have been analysed to have as little as a 0.28\% compression rate \parencite{yang2023compression}. This would have a negligible effect on file size and accounting for such a minor detail was deemed to be outside of the scope for these experiments. These experiments used the ONE simulator's \texttt{MessageEventGenerator} class \parencite{javadocs} for creating the messages during the simulation. As it takes in a range of values for determining the size of messages, the message size parameter was given a range of values in each iteration. Iteration one assumed that only the PSD graphic was transmitted, iteration two assumed that only the HVSR graphic was transmitted, and iteration three assumed both were transmitted. Therefore, the message size parameters were 600Kb - 700Kb, 1.6Mb - 1.7Mb, and 2.2Mb - 2.4Mb for iterations one, two, and three respectively. Note that the hyphen represents a range of values and not a negation operator. To ensure that all parameters in the experiments were feasible, the buffer size of the car nodes for each iteration was chosen by using intuition with the associated message size. For example, iteration one used a buffer size of 7Mb to allow each car to store a maximum of ten messages at any given time. For this reason, iterations two and three used buffer sizes of 17Mb and 24Mb respectively. Other sets of experiments used a message size of 2.2Mb - 2.4Mb and a buffer size of 24Mb by default.\\

The final set of experiments focused on node density. The parameter corresponding to the number of nodes was modelled with the assumption that each node represented 100 cars. The first iteration was designed to represent a quieter time at night when fewer residents would be travelling; since it's unrealistic to assume that tsunamis will only strike during the day. Therefore, the first iteration used 50 nodes to represent 5,000 cars. The second iteration tested the pseudo-realistic calculations in \ref{eq1} of 23,756 cars, which it represented with 238 nodes. The third iteration represented busier times, like when most people are commuting to or from work, With as many as 400 nodes representing 40,000 cars. Other sets of experiments used the pseudo-realistic figure of 238 nodes by default.\\ 

The experiments were set up to document varying results across four types of report provided by the ONE simulator \parencite{github_repo}. \texttt{MessageStatsReport} provided data for latency and delivery probability. \texttt{DeliveredMessagesReport} provided data on delivery times and hop counts. \texttt{MessageDelayReport} provided data on the delay each message experienced and the changes in delivery probability as more messages were successfully delivered. \texttt{BufferOccupancyReport} recorded the average buffer occupancy across all nodes as a percentage, alongside the variance for each average.

\section{Analysis and Critical Discussion of Experiments}
\subsection{Transmission Interface Experiments}

\begin{figure}[h!]
    \centering
    \includegraphics[width=1.0\linewidth]{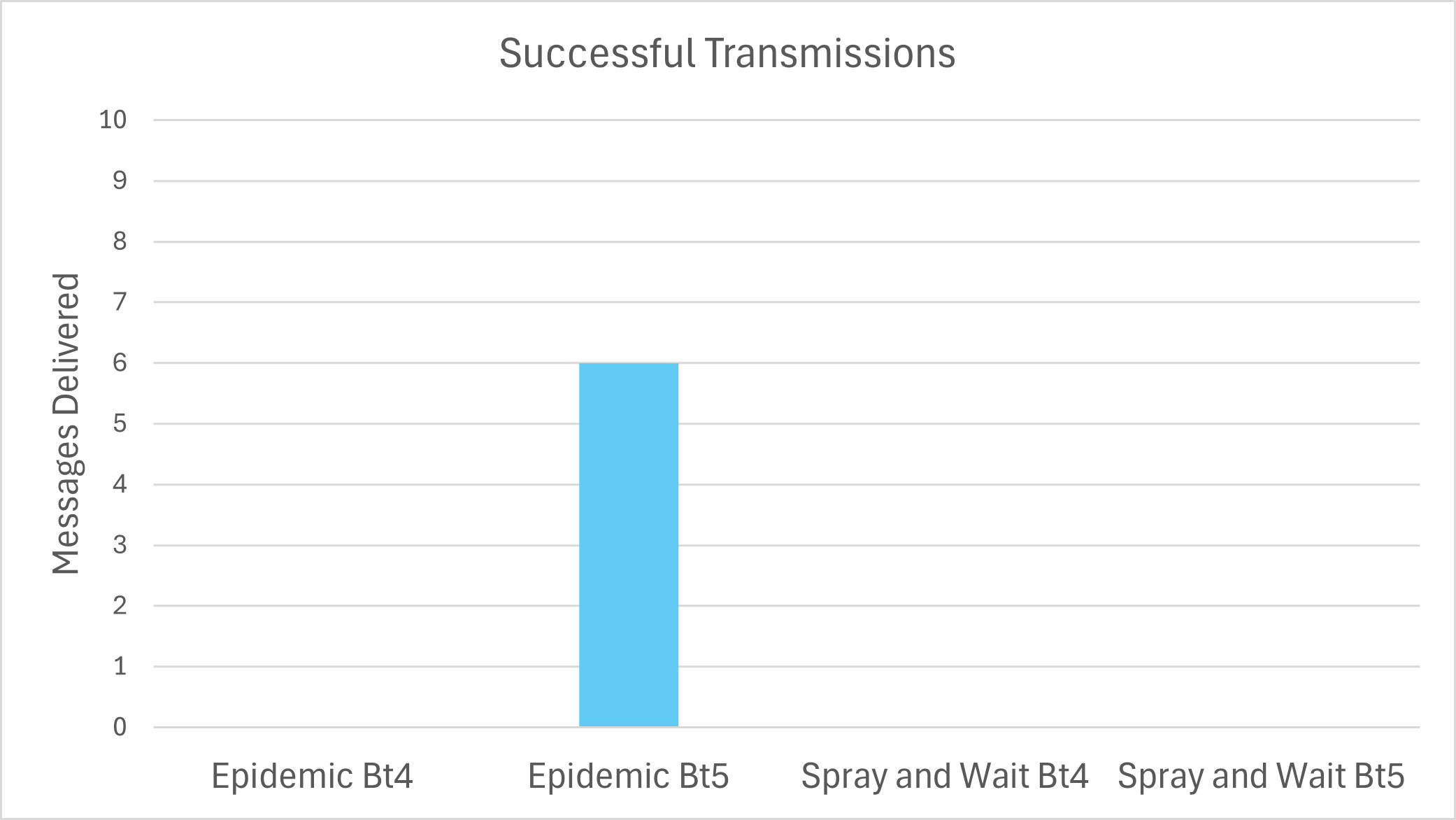}
    \caption{}
    \label{fig:bt-successful-transmissions}
\end{figure}

Figure \ref{fig:bt-successful-transmissions} shows that only epidemic with Bluetooth version 5 was able to successfully deliver any messages, and yet it still delivered very few messages. Epidemic's poor performance could signify that the nodes struggled to communicate since epidemic relies on frequent communication to rapidly spread messages. Additionally, spray and wait's inability to deliver any messages may be a sign of the environment being too big since none the \textit{L-1} nodes that received the source's messages were able to directly transmit the message to the destination. This was likely the case since the message time to live (TTL) was set to match the simulation time (30 minutes) so it was not possible for messages to expire. Neither protocol performed sufficiently for the proposed scenario since messages must be delivered in order for the BPBD centre to warn and evacuate people from the tsunami. Therefore, these protocols may be unsuitable for the proposed scenario until more powerful transmission technology is available.

\begin{figure}[h!]
    \centering
    \includegraphics[width=1.0\linewidth]{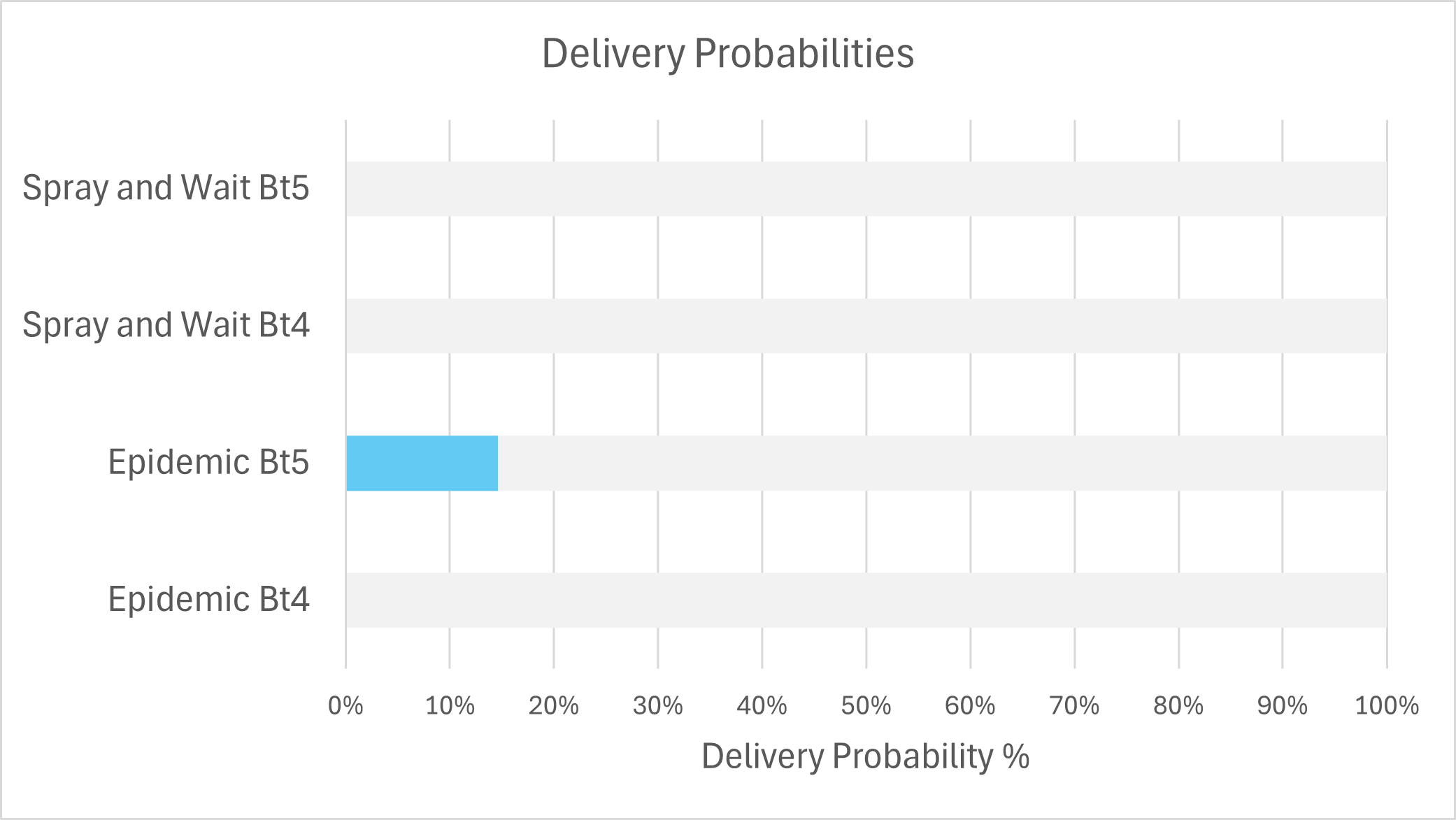}
    \caption{}
    \label{fig:bt-delivery-probabilities}
\end{figure}

The delivery success rates shown in figure \ref{fig:bt-delivery-probabilities} indicate that very few, if any, of the messages sent by the seismic station managed to reach the BPBD centre. These statistics are useful since they prove that the rate at which the source generated and spread messages wasn't the problem; if it was then the delivery probabilities would've been high. A more likely problem was that messages may have been taking too long to reach the destination in respect to the very limited simulation time. For example, the messages that did successfully reach the destination took between 20 to 40 hops between nodes. While this hop count alone indicates positive performance, since the worst case still required fewer than 20\% of the total nodes (238), it may also reveal an underlying problem that hops could be too infrequent.

\begin{figure}[h!]
    \centering
    \includegraphics[width=1.0\linewidth]{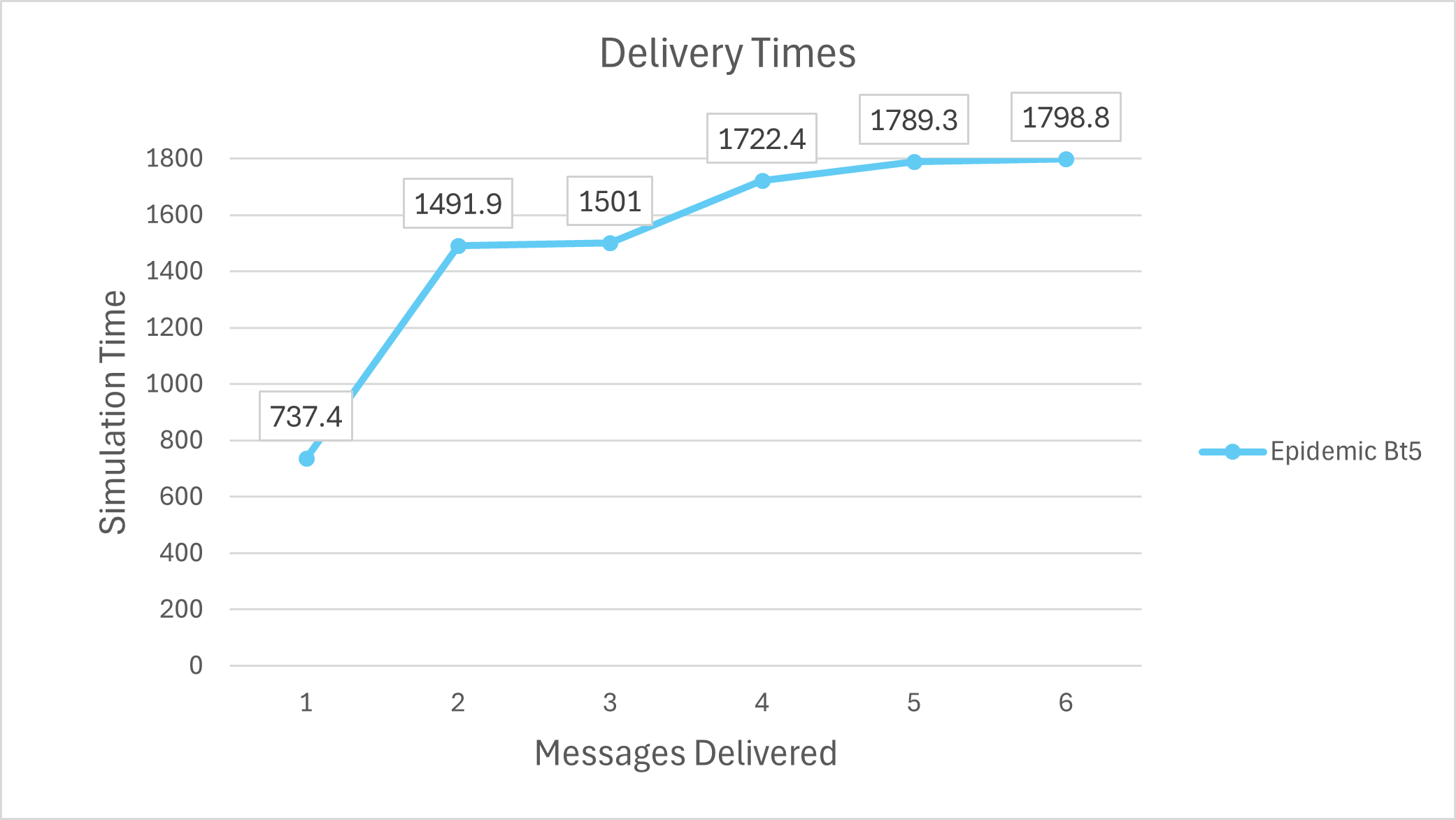}
    \caption{}
    \label{fig:bt-delivery-times}
\end{figure}

Figure \ref{fig:bt-delivery-times}'s results demonstrate that many of the messages realistically arrived far too late. From the second message delivery onwards, the messages arrived with only a few hundred seconds before the tsunami would be expected to arrive. This would give BPBD office workers at most five minutes to warn and safely evacuate everyone from the city, which is infeasible. Furthermore, attempting to evacuate hundreds of thousands of people \parencite{PaluCityCentralStatisticsAgency2023} in such a short period of time would create an additional risk of people being injured in the midst of widespread panic and rushing. Therefore, alternative protocols that can achieve lower latencies, like MaxProp \parencite{alaoui2015performance}, would likely be more suitable than epidemic and spray and wait for the proposed scenario.

\begin{figure}[h!]
    \centering
    \includegraphics[width=1.0\linewidth]{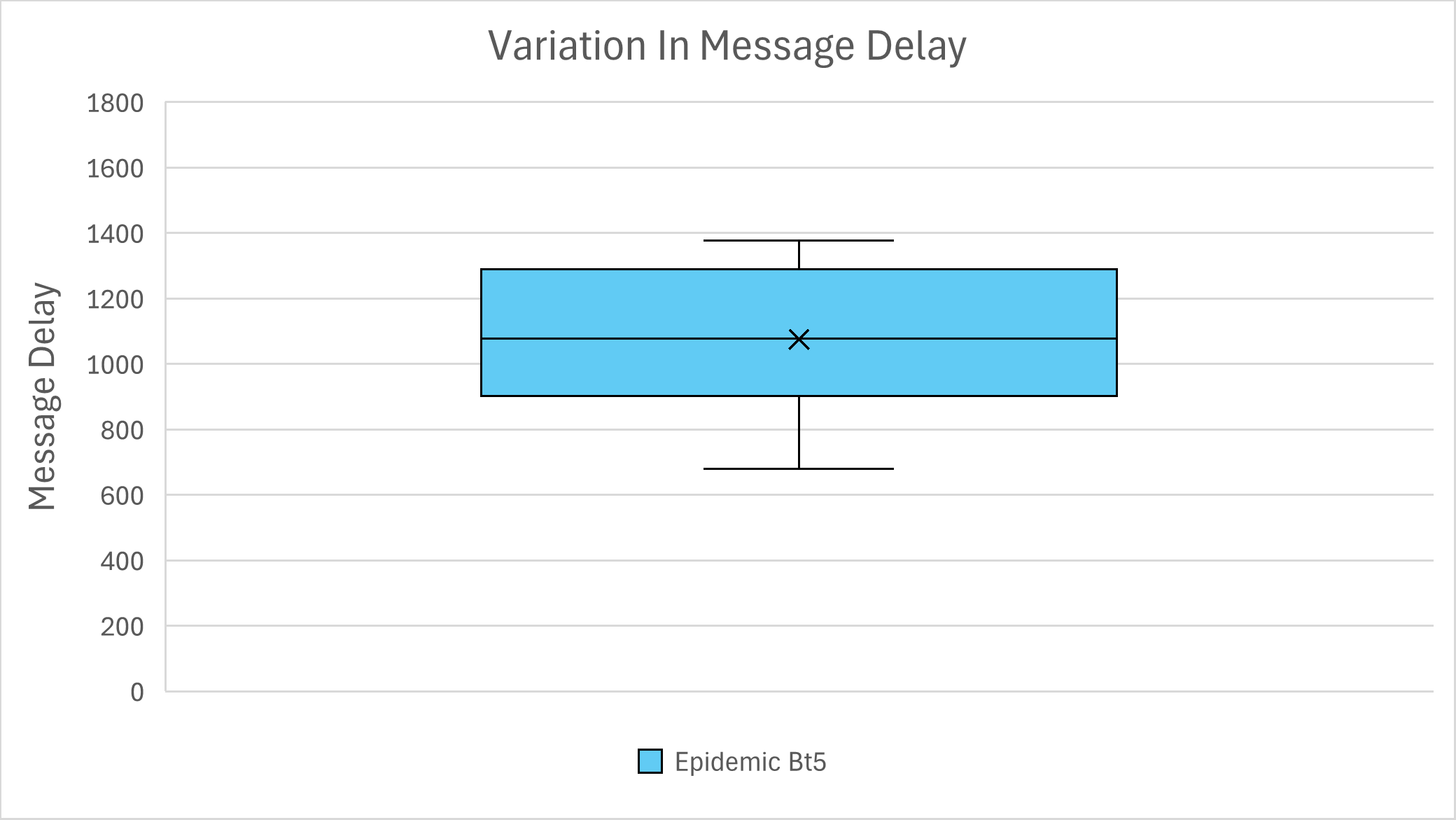}
    \caption{}
    \label{fig:bt-message-delay}
\end{figure}

Figure \ref{fig:bt-message-delay} shows a noticeable level of variance across the delays that the messages experienced. The minimum and maximum values have a fairly large difference in delay, but the interquartile range shows a reasonable range of delays. This demonstrates a relatively consistent performance which is a sign of positive reliability. While the message delays weren't as detrimental as the previously mentioned delivery times, an average delay of approximately 1,100 seconds is still too high since any messages that weren't sent in the first half of the simulation would be unlikely to have enough time to reach the destination. This could become problematic since it could prevent subsequent messages from reaching the destination which would prevent the BPBD office from receiving updates about the seismic activity. Consequently, false positive reports from the seismic station could result in the BPBD office taking unnecessary action as it may not receive the corrected follow-up messages in time. Such high message delays are unexpected when using epidemic and may arise from a lack of communication opportunities between nodes due to the complex layout of the Palu city map (figure \ref{fig:palu-in-one}) which has a lot of dead ends and corners.

\subsection{Protocol Efficiency Experiments}

\begin{figure}[h!]
    \centering
    \includegraphics[width=1.0\linewidth]{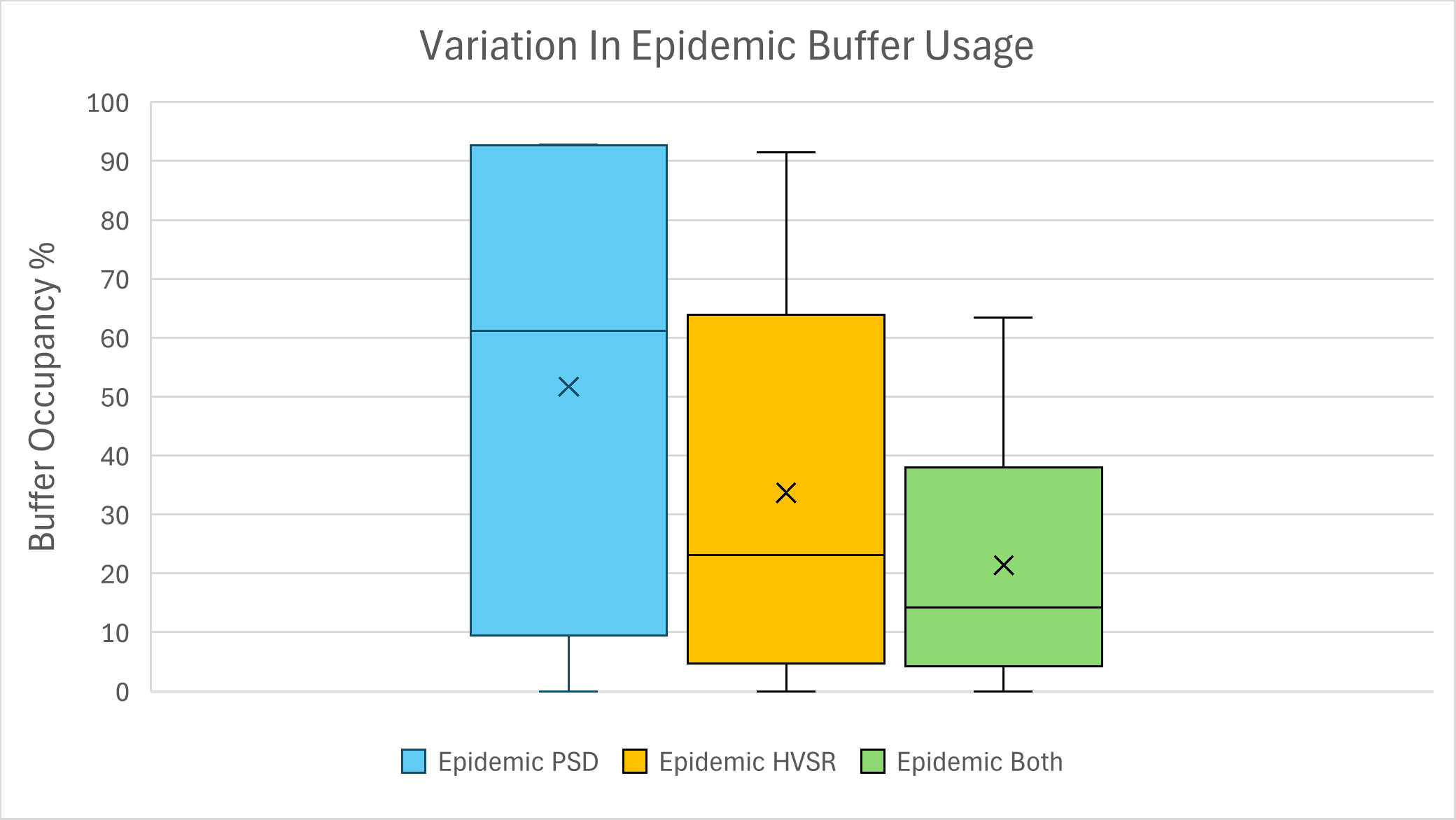}
    \caption{}
    \label{fig:epidemic-buffer-variation}
\end{figure}

Figure \ref{fig:epidemic-buffer-variation} shows some unexpected results, with the most resource expensive case, of having a message contain both the PSD and HVSR images, performing objectively better than the other two configurations that both had a smaller message size. Furthermore, the buffer size associated with each message type was designed to hold up to ten copies, so the larger buffer size for the case of both messages was not expected to advantage it in any way. Despite using more buffer space, messages with only PSD and messages with only HVSR still have surprisingly low median and mean values for simulations with epidemic routing, which often risks occupying high amounts of buffer space \parencite{Burleigh2020}. These results are very positive since it means that, on average, the node buffers should have more than enough room for multiple subsequent messages from the BPBD office, which would help with evacuation efforts. Additionally, real-world applications would use both of the images in the messages so it's promising to see that configuration providing the best and most consistent results - as denoted by its interquartile range. Despite how beneficial this behaviour is, it seems too good to be true and is likely only happening due to the poor performance of epidemic in this scenario. Since the nodes are rarely communicating, they will not share messages very often and so therefore nodes will be storing fewer messages on average. Consequently, this behaviour is probably unrealistic and unrepresentative of how epidemic may perform in a scenario with less restrictive parameters.\\

\begin{figure}[h!]
    \centering
    \includegraphics[width=1.0\linewidth]{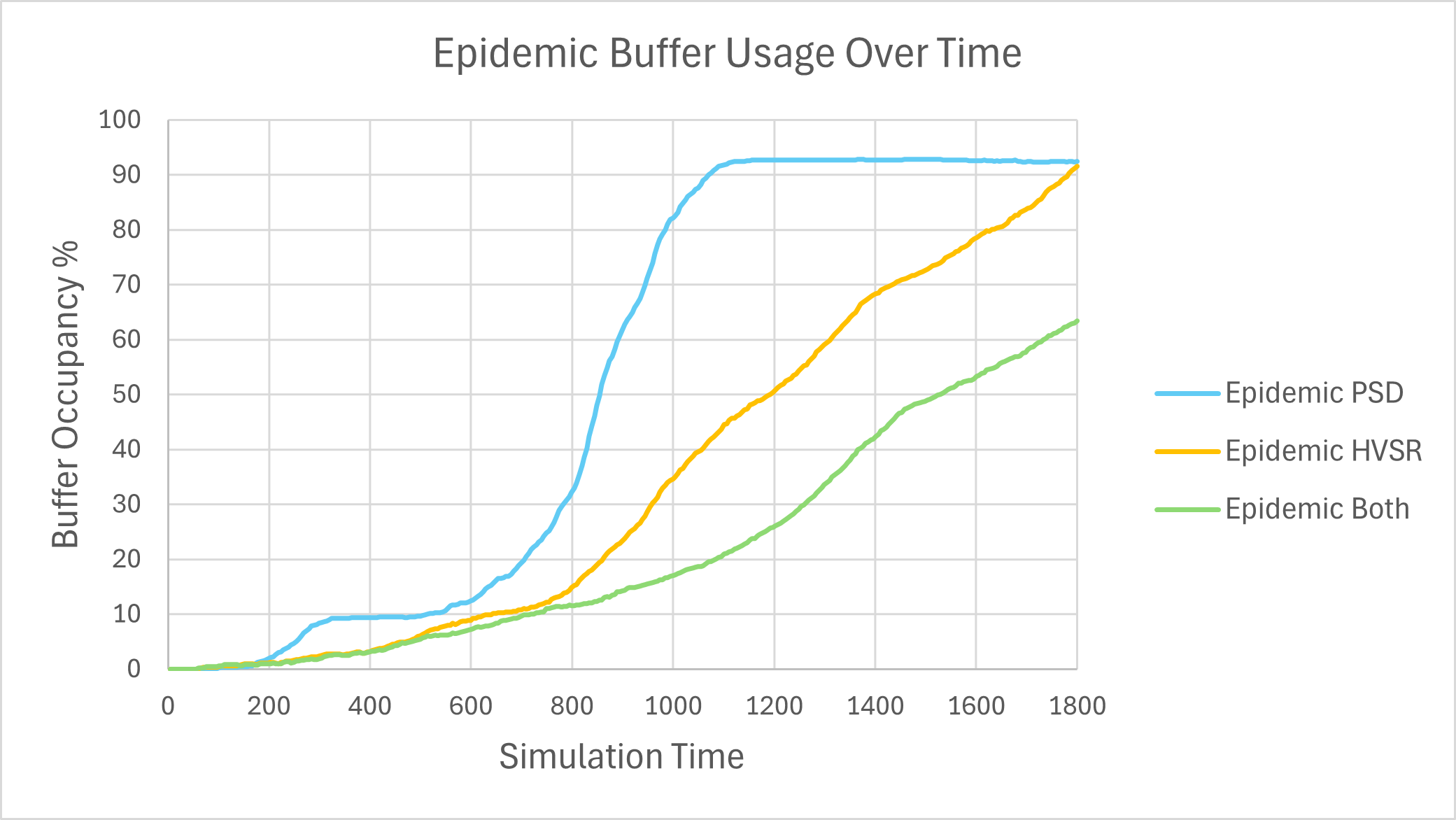}
    \caption{}
    \label{fig:epidemic-buffer-ot}
\end{figure}

From figure \ref{fig:epidemic-buffer-ot} it can be inferred that messages only containing PSD images filled up buffers very quickly at approximately halfway through the simulation time. Messages containing only HVSR and messages containing both had a much smoother increase, with the latter being more gradual and ending up with a lower buffer occupancy at the end of the simulation. These results suggest that buffers with smaller sizes filled up much quicker, which seems trivial but becomes less clear when factoring in that the buffer sizes were designed in proportion to the message sizes. While the reason for this is unclear, it can be concluded that smaller buffers filled faster than larger buffers irregardless of message size, which would emphasise the importance of having large enough buffers in this scenario.

\begin{figure}[h!]
    \centering
    \includegraphics[width=1.0\linewidth]{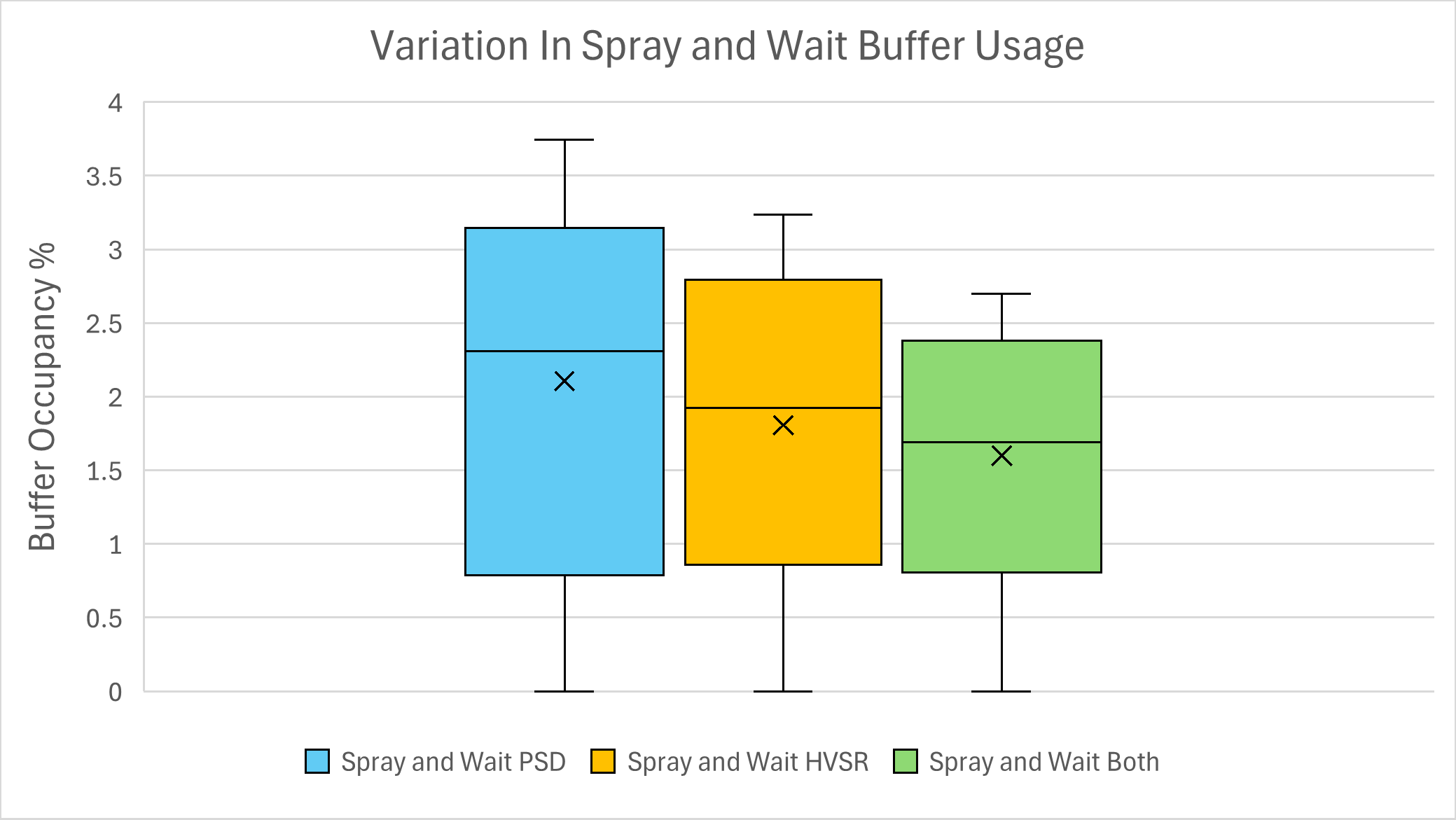}
    \caption{}
    \label{fig:snw-buffer-variation}
\end{figure}

The results displayed in figure \ref{fig:snw-buffer-variation} suggest that, like with epidemic, messages containing both images objectively required the least amount of buffer space. Although, all three message configurations performed very similarly so the difference between the best and worst performing was minimal. More importantly, all message configurations used less than 4\% of the available buffer space. While this may seem beneficial, the nodes had significant amounts of unnecessary buffer space in the best case. In the worst - and by far most probable - case, the negligible buffer space usage would've been due to barely any messages being spread during the spray phase. This may well have been due to the maximum number of copies \textit{L} of each message being too low or messages being sent too infrequently. These would be easy to fix and wouldn't conflict with the restrictions imposed by the scenario. However, it could also be due to very few nodes being close enough to the seismic station when it attempted to spray messages. Previously explored results, like in figures \ref{fig:bt-successful-transmissions} and \ref{fig:bt-message-delay}, showed that communication between nodes didn't occur often enough and this could certainly have included communication between vehicle nodes and the seismic station. Since the messages are sprayed to the first \textit{L-1} nodes that the source encounters \parencite{Burleigh2020}, if the source does not encounter \textit{L-1} nodes before attempting to spray new messages then it will always spray fewer messages than intended. This could be a clear explanation as to why the buffer usage is so low and why spray and wait was unable to deliver any messages, as discussed earlier. Behaviour like this would be unacceptable for the scenario since the seismic station's readings would be rendered almost pointless if it wasn't able to share them effectively due to the network. Furthermore, this could threaten many lives as the population may be completely unaware of a fast approaching tsunami.\\\\\\\\\\\\

\begin{figure}[h!]
    \centering
    \includegraphics[width=1.0\linewidth]{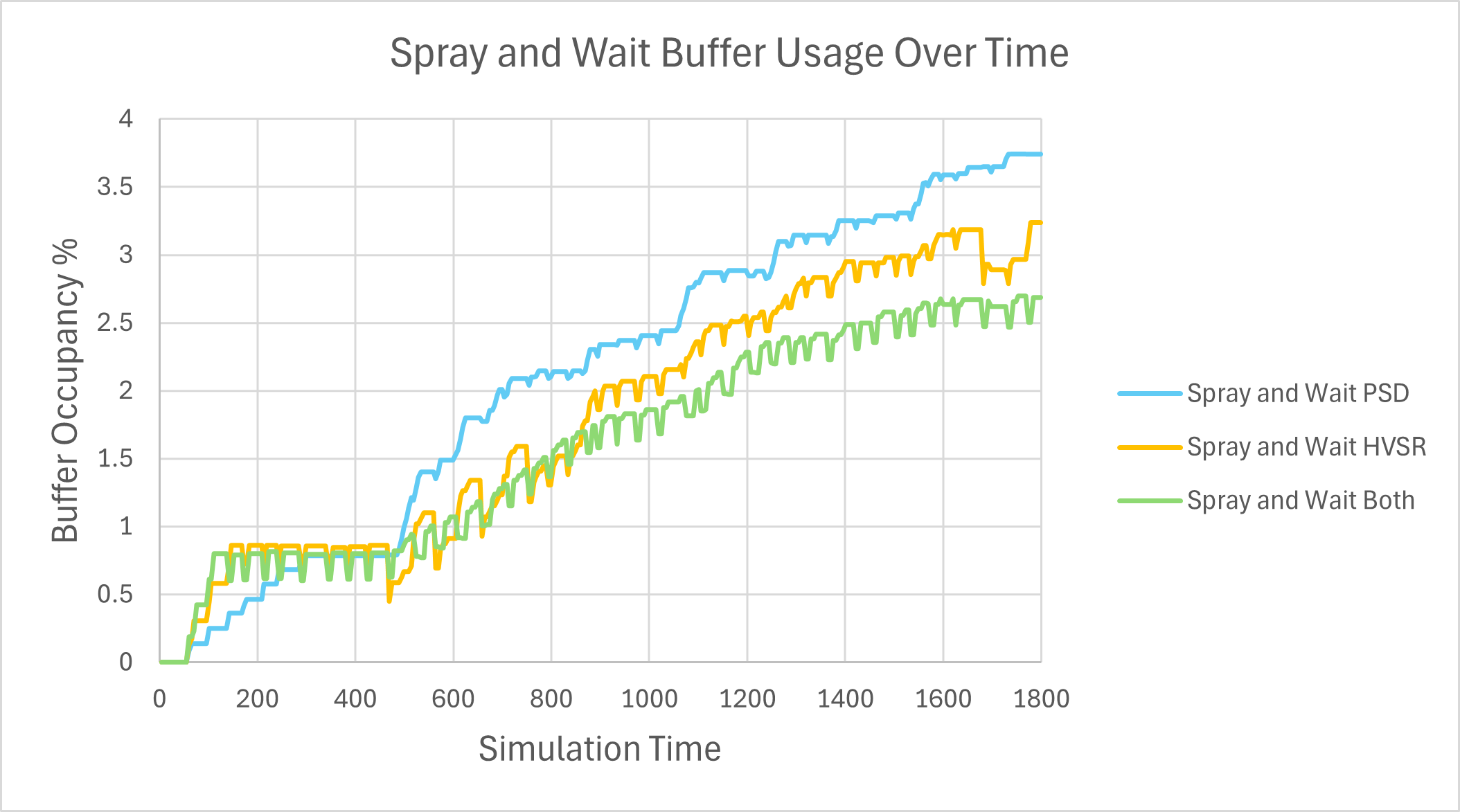}
    \caption{}
    \label{fig:snw-buffer-ot}
\end{figure}

Figure \ref{fig:snw-buffer-ot} indicates that there was not much difference between how the three message configurations performed on average; unlike epidemic (figure \ref{fig:epidemic-buffer-ot}) which had a much starker difference in performance. The frequent dips were intriguing and likely a result of the amount of variance in the average buffer usage measures constantly increasing and decreasing. The average buffer occupancy for epidemic (figure \ref{fig:epidemic-buffer-ot}) did not experience such large jumps in variance and is therefore much smoother. The consistent performances are positive and show more reasonable behaviours, since it was expected that buffer sizes wouldn't produce broadly different results as they were all proportional to their message sizes. However, this similar behaviour is insignificant since all three message configurations yielded unrealistic, insufficient buffer usage rates that render none of them applicable for the scenario.

\subsection{Node Density Experiments}

\begin{figure}[h!]
    \centering
    \includegraphics[width=1.0\linewidth]{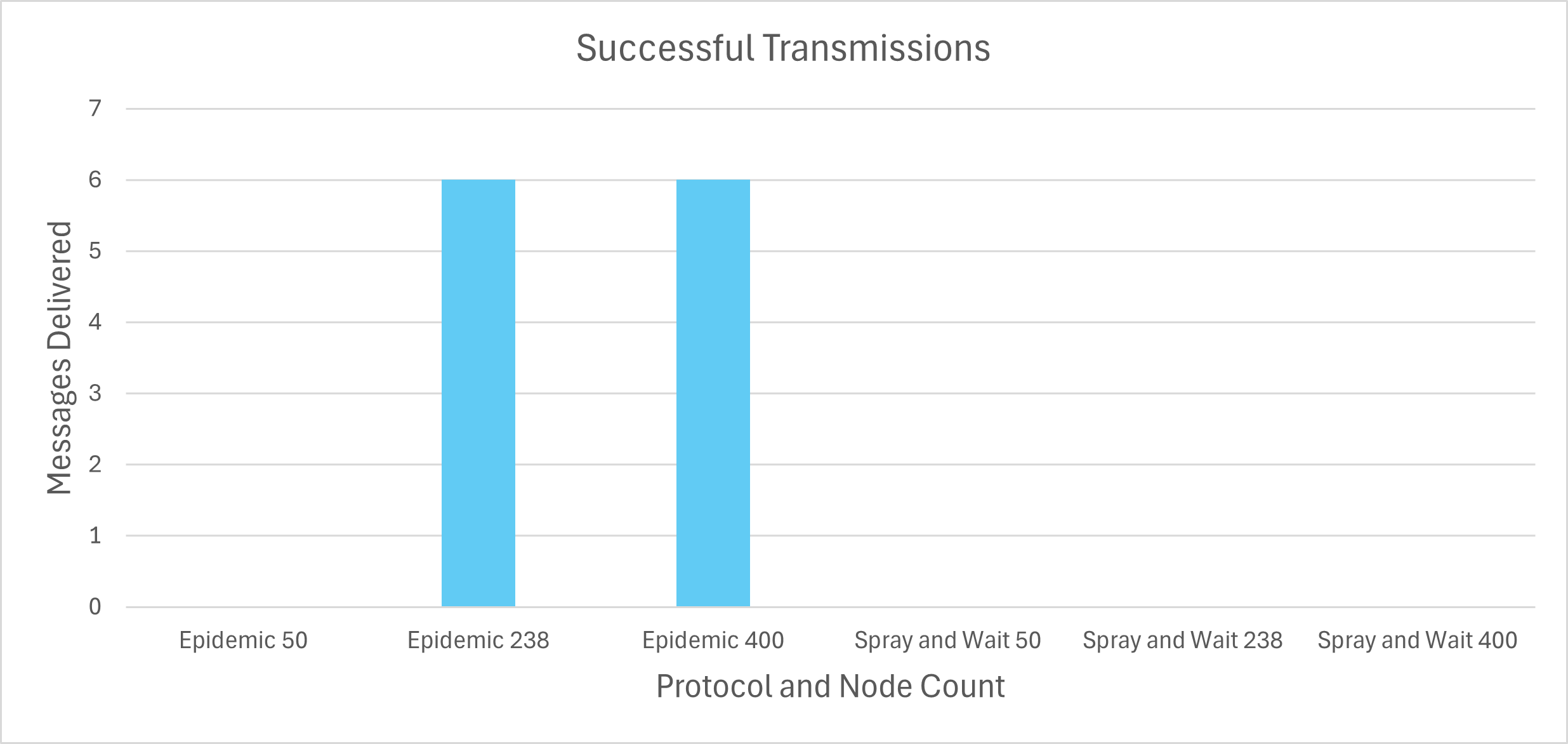}
    \caption{}
    \label{fig:node-successful-transmissions}
\end{figure}

Figure \ref{fig:node-successful-transmissions} shows that increasing the number of cars travelling around Palu city didn't meaningfully improve the performance of epidemic nor spray and wait. This provides useful insight, as it eliminates the possibility of both protocols' poor performance being attributed to a lack of nodes. However, this behaviour is strange since increasing the number of nodes would be expected to also increase the number of communication opportunities. It is unlikely that the frequency of communication increased with more nodes, beyond 238, since the configuration with 400 nodes had a very similar performance. Although, this is a small sample size and adding more nodes may result in improved performance; but doing so would make the simulation far too unrealistic and yield unhelpful results. As previously mentioned, the communication between nodes being too uncommon may have been a result of the Palu city environment (figure \ref{fig:palu-in-one}) being too large and complex.

\begin{figure}[h!]
    \centering
    \includegraphics[width=1.0\linewidth]{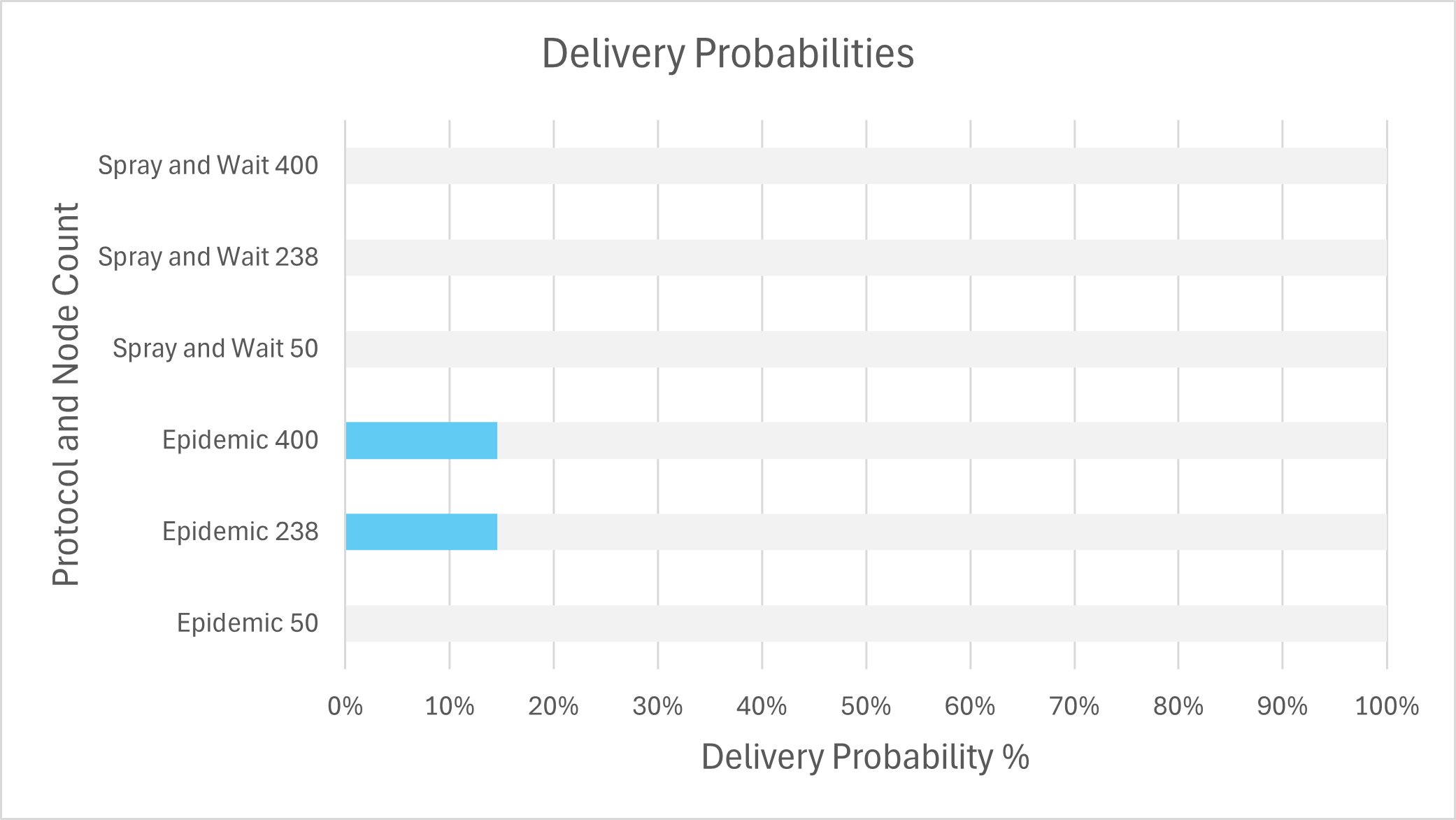}
    \caption{}
    \label{fig:node-delivery-probabilities}
\end{figure}

Similarly to the previous figure, figure \ref{fig:node-delivery-probabilities} demonstrates that increasing the number of nodes did not make the DTN any more suitable for the proposed scenario. These delivery probabilities are far too low and would render a DTN with epidemic or spray and wait as a completely unviable option. A possible approach to make such low delivery rates work would be to send a large number of packets to ensure that a reasonable amount of packets are still successfully delivered. However, this approach would not be feasible since it would fill up node buffers too quickly, especially when using epidemic, and attempting to continuously transfer such a large amount of data risks crippling the network as a whole.\\

\begin{figure}[h!]
    \centering
    \includegraphics[width=1.0\linewidth]{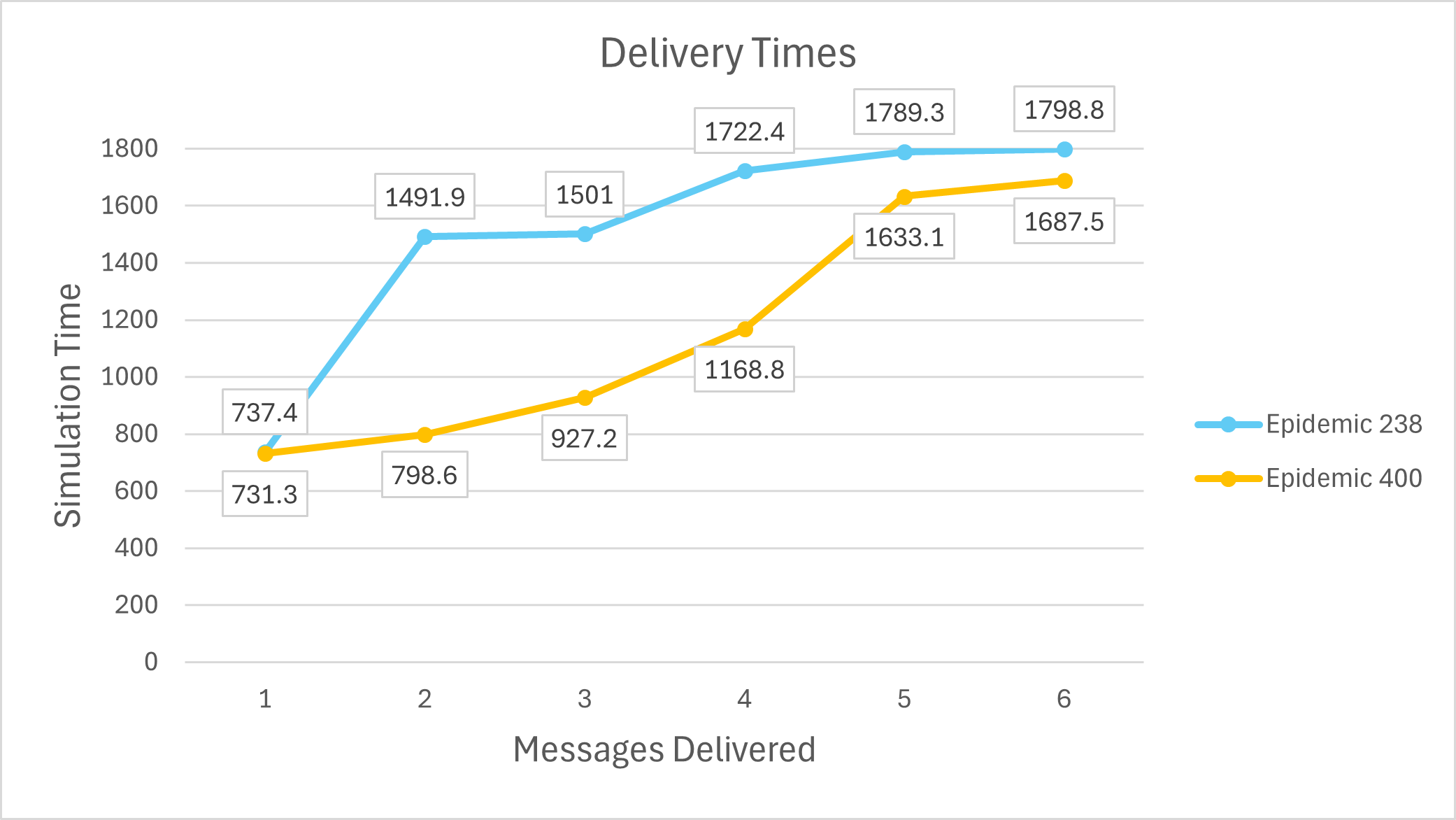}
    \caption{}
    \label{fig:node-delivery-times}
\end{figure}

Figure \ref{fig:node-delivery-times} suggests that, despite what the results of the previous two figures showed, increasing the number of nodes does improve the performance of epidemic to some extent. Using 400 nodes would allow the seismic station to successfully deliver multiple warning/informational messages to the BPBD centre in a satisfactory duration of time. However, it must be considered that the case of having 400 vehicle nodes was modelling a real-world scenario with many people commuting (see section \ref{section-4}), which would likely have highly congested roads. Therefore, evacuation efforts could become much more difficult and time-consuming since traffic would prevent people from quickly escaping in vehicles. Alongside this, a large number of cars could block roads and potential evacuation routes off from people fleeing on foot \parencite{car_evacuation_feasibility}. Therefore, the delivery times achieved by adding more nodes may still be insufficient for the scenario.\\

\begin{figure}[h!]
    \centering
    \includegraphics[width=1.0\linewidth]{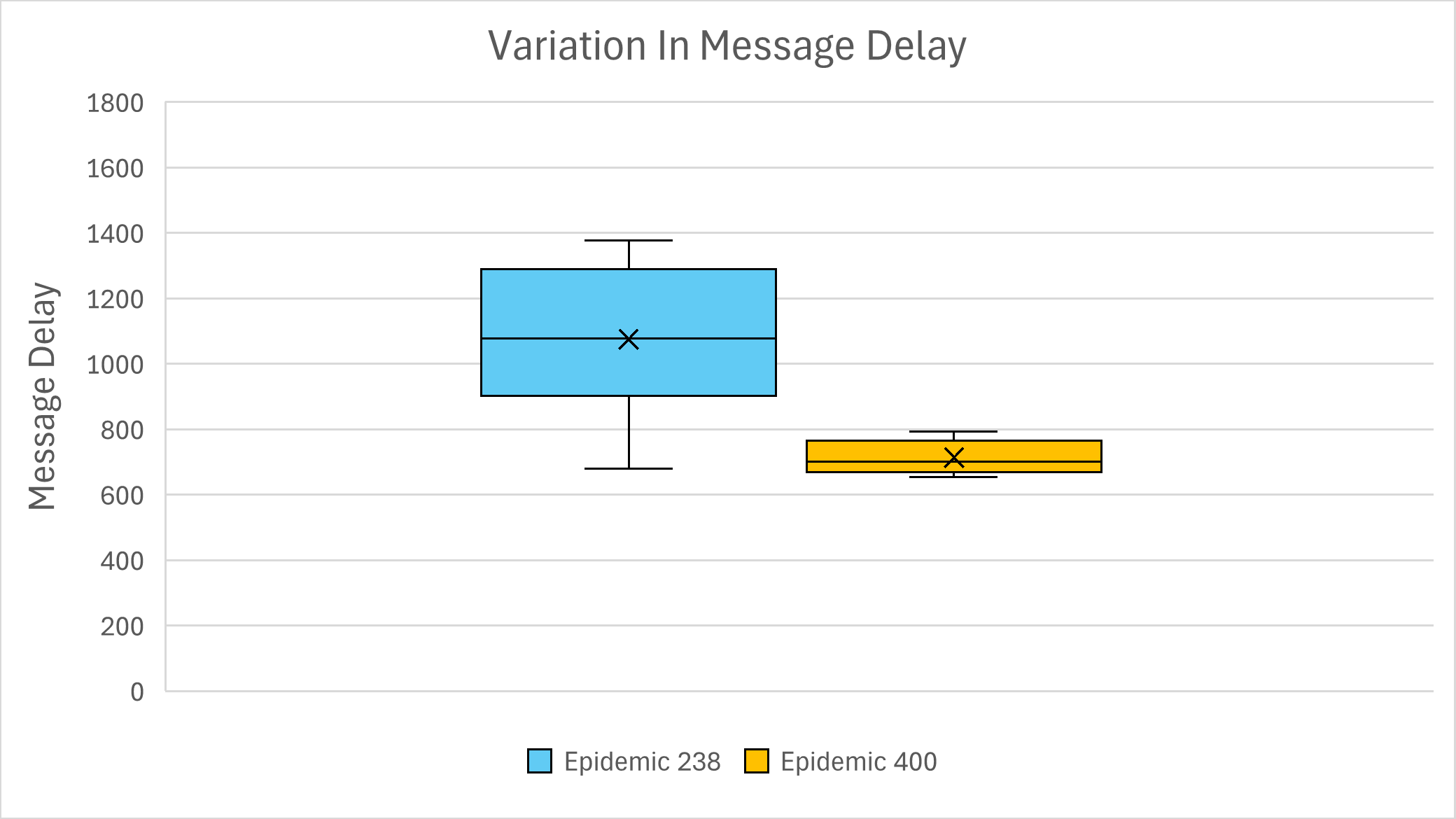}
    \caption{}
    \label{fig:node-message-delay}
\end{figure}

The variations shown in figure \ref{fig:node-message-delay} indicate that adding more nodes reduced the variance in message delay and made it more consistent. This makes sense since increasing the number of nodes in a network will decrease its sparsity. This reduction in variance is important as minimising variance helps to improve reliability which is crucial in a scenario where people's lives can rely on how quickly a warning message is sent after receiving signs of a tsunami. However, these results also show that the expected number of cars in Palu (represented by 238 nodes) would produce suboptimal amounts of variance, which would hinder the applicability of a DTN with epidemic for this scenario. Furthermore, the results in figure \ref{fig:snw-buffer-ot} showed that spray and wait produced much worse variance and would certainly not be reliable enough.\\

\section{Conclusion}

Overall, based on the results of the experiments and the nature of this paper's scenario, the proposed DTN approach would not be suitable for routing messages between a seismic station and a local BPBD office. This is primarily because neither epidemic nor spray and wait could successfully deliver enough messages before the tsunami was expected to arrive. Additionally, the messages that did arrive often experienced too much delay or were too unreliable given how much variance they had in their delay. Despite these shortcomings, other routing protocols, like MaxProp \parencite{maxprop} and PRoPHET \parencite{prophet}, may be sufficient; especially since previous research found that MaxProp displayed far superior performance in comparison to epidemic and spray and wait \parencite{alaoui2015performance, gamit2014evaluation}. If other DTN configurations were able to successfully and reliably deliver messages in time then they could be highly beneficial in helping evacuation efforts. The ability to still deliver messages despite disruptions in connections \parencite{Vasilakos2025-zk} is valuable since many tsunamis are caused by earthquakes \parencite{MAKINOSHIMA2020100113} which may hinder or destroy the infrastructure that a populated region relies on for constant communication. In these cases, traditional TCP/IP communication would be unusable since it relies on a constant, end-to-end path \parencite{Vasilakos2025-zk}. DTNs could therefore be used as backup systems for warnings and rescue efforts in a variety of natural disaster situations.\\

\section{Related Works}
Some related works include \parencite{kawano2025delaytolerantnetworkingtsunamievacuation} which found success in using DTNs for tsunami evacuation in Hachijojima, Japan. It found the most success with epidemic routing, which was able to deliver approximately 15 emergency messages in the 30 minute window allocated for the paper's scenario. However, it differs from this paper since its evacuation scenario assumed that the population had already been warned of the tsunami ahead of the 30 minute evacuation window. The difference in results between the two papers could indicate that DTNs may be better suited for use with natural disasters that have already been identified, rather than as early warning systems.\\

\parencite{dtn_for_evacuation} explores the application of DTNs for providing people with information on effective evacuation routes during an emergency scenario. It found that DTNs succeeded in helping people evacuate approximately 12\% faster, which would work well in conjunction with the approach that this paper is providing.\\ 

Many other related works investigated the suitability of DTNs for post-disaster scenarios. For example, \parencite{uav_dtns} found that Unmanned Aerial Vehicles (UAVs) could be used as mobile nodes in a DTN architecture to support rescue teams in communicating with each other while searching for people and providing supplies in natural disaster scenarios caused by heavy rainfall. It found that as few as two UAVs were able to sufficiently route messages between rescue teams. \parencite{Bhattacharjee2015} investigated how DTNs could be used in post-disaster relief based on the flooding in Uttarakhand in 2013. It treated relief camp volunteers with mobile phones as DTN nodes and placed additional emphasis on energy efficiency. The results were positive and found that DTNs could be effectively used in the aftermath of natural disasters to provide communication for rescuing people and providing relief.\\

\parencite{novel_framework_large_scale_emergency_communications} proposes the Energy Efficient CongEstion Aware cross-layer intelligent
Emergency Framework (E$^3$F) which is designed to work in large-scale disaster scenarios where traditional communication can fail due congestion and limited power supplies. It works through using three types of nodes: resource deprived nodes representing individual people within the areas of emergency, nodes that represent fire engines that can replicate the resource deprived node's messages alongside being able to send and respond to queries, and external helper nodes that operate outside of the emergency area to improve the connectivity and available bandwidth. Experiments involved the use of E$^3$F and MaxProp, with the results showing that E$^3$F consumed half as much energy as MaxProp for data dissemination and answering queries while yielding a higher delivery success rate than energy constrained MaxProp when using a larger number of external helper nodes. These are promising results given the risk that little energy may be available in future emergency scenarios and that they were able to handle emergency situations on a larger geographical scale.\\

\section{Future Work}
Most of these existing works provide promising solutions for saving lives after a natural disaster has occurred, but all of them focus on reactive action. Their approaches do not offer help before the natural disasters occur, which forfeits an opportunity to save more lives by setting up the means of communication for proactive efforts. This paper proposed a more proactive approach that aims to save lives by taking action before the tsunami arrives. However, this approach could not sufficiently operate within the restricted time frame. Therefore, future work will focus on applying DTNs to the earthquakes that may precede a tsunami in order to provide more time for evacuation and make post-tsunami relief easier.\\

While the ONE simulator is helpful for conducting experiments with new networking approaches, it's limited in the extent to which it can represent real-world conditions. Given the life-critical nature of the scenarios which this paper's proposed and future works address, it's important to replicate real-world conditions as closely as possible so that the feasibility of such approaches can be assessed as accurately as possible. The Mobile Opportunistic and Disconnection Tolerant Networking (MODiToNeS) \parencite{towards_low_cost} platform could be used in conjunction with ONE to help achieve this. Information about Palu's, or any other region covered in future works, network topology could be passed into MODiToNeS to conduct experiments on accurate representations of such topologies since MODiToNeS is able to replicate them using topological information \parencite{towards_low_cost}. Furthermore, MODiToNeS' can provide real-time analytics and predictions \parencite{towards_low_cost} that could allow experiments for future works to yield results that better indicate their feasibility for deployment in real world scenarios.\\

However, MODiToNeS needs to be deployed in the real world, and this could be done with multiple heterogeneous nodes attached to vehicles or drones, as showcased in \parencite{enabling_real_time_communications} as part of their deployments in agricultural and urban environments. Future works based on this paper could be deployed in a similar way with multiple MODiToNeS nodes, each attached to a different car, to represent vehicular nodes. These cars could drive around a small area within Palu, or wherever else future works are taking place, such that each car takes a different loop/route around the chosen area. The results of which could then be evaluated using a variety of metrics like those evaluated earlier in this paper. Since the experiments would realistically only be able to use a few cars in a small area of the city, to ensure that the cars have a reasonable chance to contact, the results would need to be extrapolated in order to represent the entire area that the future works would aim to operate in. These extrapolated results may then provide fairly reasonable insight into the feasibility of the future works' approaches.\\

Approaches in future works may account for resource constraints within the emergency environments since damage to networking infrastructure, like what may have happened in Palu, can limit the available resources. \parencite{adaptive_real_time_predictive} proposes an adaptive real-time predictive multi-layer caching and forwarding approach called CafRepCache, which can handle resource restrictions and dynamically changing resources \parencite{adaptive_real_time_predictive}. CafRepCache uses a resource management module consisting of predictive buffer storage, smart scheduling, and packet prioritisation; in addition to using predictive heuristics that use information on the local available resources \parencite{adaptive_real_time_predictive}. These features proved beneficial as CafRepCache achieved the highest success ratio, the lowest delays, and typically the least packet loss for content retrieval when compared to leading and benchmark protocols \parencite{adaptive_real_time_predictive}. Given the environments and scenarios that this paper's future works aim to cover, adopting similar resource management and leverage techniques could make future approaches far more robust to limited network resources, which can be expected in disaster-struck environments; all while achieving sufficiently low latency and high success rates for such approaches to be feasible for real-world deployment.\\

\printbibliography[title=Bibliography]

@book{basagni2004mobile,
  title={Mobile Ad Hoc Networking},
  author={Basagni, S. and Conti, M. and Giordano, S. and Stojmenovic, I.},
  isbn={9780471656883},
  lccn={2004062093},
  url={https://books.google.co.uk/books?id=GnkcHEsxAigC},
  year={2004},
  publisher={Wiley}
}

@article{hoebeke2004overview,
  title={An overview of mobile ad hoc networks: applications and challenges},
  author={Hoebeke, Jeroen and Moerman, Ingrid and Dhoedt, Bart and Demeester, Piet},
  journal={Journal-Communications Network},
  volume={3},
  number={3},
  pages={60--66},
  year={2004},
  publisher={THE COMMUNICATIONS NETWORK}
}

@article{ABOLHASAN20041,
title = {A review of routing protocols for mobile ad hoc networks},
journal = {Ad Hoc Networks},
volume = {2},
number = {1},
pages = {1-22},
year = {2004},
issn = {1570-8705},
doi = {https://doi.org/10.1016/S1570-8705(03)00043-X},
url = {https://www.sciencedirect.com/science/article/pii/S157087050300043X},
author = {Mehran Abolhasan and Tadeusz Wysocki and Eryk Dutkiewicz}
}

@BOOK{Vasilakos2025-zk,
  title     = "Delay Tolerant Networks : Protocols and Applications",
  author    = "Vasilakos, A V and Zhang, Y and Spyropoulos, T",
  publisher = "Taylor \& Francis Group",
  year      =  2011,
  address   = "Milton. Ebook Central"
}

@article{CHLAMTAC200313,
title = {Mobile ad hoc networking: imperatives and challenges},
journal = {Ad Hoc Networks},
volume = {1},
number = {1},
pages = {13-64},
year = {2003},
issn = {1570-8705},
doi = {https://doi.org/10.1016/S1570-8705(03)00013-1},
url = {https://www.sciencedirect.com/science/article/pii/S1570870503000131},
author = {Imrich Chlamtac and Marco Conti and Jennifer J.-N. Liu}
}

@INPROCEEDINGS{8226101,
  author={Ukey, Nimish and Kulkarni, Lalit},
  booktitle={2017 2nd International Conference for Convergence in Technology (I2CT)}, 
  title={Implementation of energy efficient algorithm in delay tolerant networks}, 
  year={2017},
  volume={},
  number={},
  pages={93-99},
  doi={10.1109/I2CT.2017.8226101}
}

@ARTICLE{e100-b_1_2,
  author={Masato TSURU and Mineo TAKAI and Shigeru KANEDA and Agussalim and Rabenirina AINA TSIORY},
  journal={IEICE TRANSACTIONS on Communications}, 
  title={Towards Practical Store-Carry-Forward Networking: Examples and Issues}, 
  year={2017},
  volume={E100-B},
  number={1},
  pages={2-10},
  doi={10.1587/transcom.2016CQI0001},
  ISSN={1745-1345},
  month={1}
}

@INPROCEEDINGS{6402866,
  author={Del Duca Almeida, Virgil and Oliveira, André B. and Macedo, Daniel F. and Nogueira, José Marcos S.},
  booktitle={2012 IFIP Wireless Days}, 
  title={Performance evaluation of MANET and DTN routing protocols}, 
  year={2012},
  volume={},
  number={},
  pages={1-6},
  doi={10.1109/WD.2012.6402866}
}

@inproceedings{10.1145/1015467.1015484,
author = {Jain, Sushant and Fall, Kevin and Patra, Rabin},
title = {Routing in a delay tolerant network},
year = {2004},
isbn = {1581138628},
publisher = {Association for Computing Machinery},
address = {New York, NY, USA},
url = {https://doi.org/10.1145/1015467.1015484},
doi = {10.1145/1015467.1015484},
booktitle = {Proceedings of the 2004 Conference on Applications, Technologies, Architectures, and Protocols for Computer Communications},
pages = {145–158},
numpages = {14},
location = {Portland, Oregon, USA},
series = {SIGCOMM '04}
}

@article{BENAMAR2014141,
title = {Routing protocols in Vehicular Delay Tolerant Networks: A comprehensive survey},
journal = {Computer Communications},
volume = {48},
pages = {141-158},
year = {2014},
note = {Opportunistic networks},
issn = {0140-3664},
doi = {https://doi.org/10.1016/j.comcom.2014.03.024},
url = {https://www.sciencedirect.com/science/article/pii/S0140366414001212},
author = {Nabil Benamar and Kamal D. Singh and Maria Benamar and Driss {El Ouadghiri} and Jean-Marie Bonnin}
}

@inproceedings{keranen-theone,
 author = {Ari Ker\"{a}nen and J\"{o}rg Ott and Teemu K\"{a}rkk\"{a}inen},
 title = {{The ONE Simulator for DTN Protocol Evaluation}},
 booktitle = {SIMUTools '09: Proceedings of the 2nd International Conference on Simulation Tools and Techniques},
 year = {2009},
 isbn = {978-963-9799-45-5},
 pages = {},
 location = {Rome, Italy},
 doi = {},
 publisher = {ICST},
 address = {New York, NY, USA},
}

@article{keranen2008opportunistic,
  title={Opportunistic network environment simulator},
  author={Keranen, Ari},
  journal={Special Assignment report, Helsinki University of Technology, Department of Communications and Networking},
  year={2008}
}

@misc{ncei_tsunami_database,
  title        = {NCEI/WDS Global Historical Tsunami Database},
  author       = {{National Geophysical Data Center / World Data Service}},
  howpublished = {NOAA National Centers for Environmental Information},
  year         = {2025},
  doi          = {10.7289/V5PN93H7},
  note         = {Accessed: 2025-11-21}
}

@article{SABAH2023474,
title = {A comprehensive report on the 28th September 2018 Indonesian Tsunami along with its causes},
journal = {Natural Hazards Research},
volume = {3},
number = {3},
pages = {474-486},
year = {2023},
issn = {2666-5921},
doi = {https://doi.org/10.1016/j.nhres.2023.06.003},
url = {https://www.sciencedirect.com/science/article/pii/S2666592123000628},
author = {Nazeel Sabah and Arjun Sil}
}

@article{palu_tsunami_insights,
author = {Cilia, Marcella and Mooney, Walter and Nugroho, Cahyo},
year = {2021},
month = {09},
pages = {},
title = {Field Insights and Analysis of the 2018 Mw 7.5 Palu, Indonesia Earthquake, Tsunami and Landslides},
volume = {178},
journal = {Pure and Applied Geophysics},
doi = {10.1007/s00024-021-02852-6}
}

@misc{BBC2018,
   author = {BBC},
   month = {10},
   title = {Indonesia earthquake and tsunami: How warning system failed the victims - BBC News},
   url = {https://www.bbc.co.uk/news/world-asia-45663054},
   year = {2018}
}

@inproceedings{amrullah2023challenges,
  title={Challenges of Local Disaster Management Agencies (BPBD) in Realizing Inclusive Disaster Management in Indonesia},
  author={Amrullah, Satrio},
  booktitle={E3S Web of Conferences},
  volume={447},
  pages={06001},
  year={2023},
  organization={EDP Sciences}
}

@Book{Burleigh2020,
author={Burleigh, Scott
and Obraczka, Katia
and da Silva, Aloizio Pereira},
title={Delay and Disruption Tolerant Networks: Interplanetary and Earth-Bound -- Architecture, Protocols, and Applications},
year={2020},
publisher={Taylor {\&} Francis Group},
note={Recommended},
isbn={9780367571146}
}

@misc{vahdat2000epidemic,
  title={Epidemic routing for partially connected ad hoc networks},
  author={Vahdat, Amin and Becker, David and others},
  year={2000},
  publisher={Technical Report CS-200006, Duke University Durham}
}

@article{geist2006tsunami,
  title={Tsunami: wave of change},
  author={Geist, Eric L and Titov, Vasily V and Synolakis, Costas E},
  journal={Scientific American},
  volume={294},
  number={1},
  pages={56--63},
  year={2006},
  publisher={JSTOR}
}

@article{gonzalez1999tsunami,
 ISSN = {00368733, 19467087},
 URL = {http://www.jstor.org/stable/26058242},
 author = {Frank I. González},
 journal = {Scientific American},
 number = {5},
 pages = {56--65},
 publisher = {Scientific American, a division of Nature America, Inc.},
 title = {TSUNAMI!},
 urldate = {2025-11-28},
 volume = {280},
 year = {1999}
}

@book{dudley1998tsunami,
  title={Tsunami!},
  author={Dudley, Walter C},
  year={1998},
  publisher={University of Hawaii Press}
}

@misc{most_destructive_tsunamis,
   title = {The most destructive Tsunamis | Palu Sulawesi, Indonesia 2018},
   url = {https://www.sms-tsunami-warning.com/pages/tsunami-palu-sulawesi-indonesia-2018}
}

@article{gamit2014evaluation,
  title={Evaluation of DTN routing protocols},
  author={Gamit, Vrunda and Patel, Hardik},
  journal={INTERNATIONAL JOURNAL OF ENGINEERING SCIENCES \& RESEARCH TECHNOLOGY},
  year={2014}
}

@article{alaoui2015performance,
  title={The performance of DTN routing protocols: a comparative study},
  author={Alaoui, EA Abdellaoui and Agoujil, Said and Hajar, Moha and Qaraai, YOUSSEF},
  journal={WSEAS Transactions on Communications},
  volume={14},
  pages={121--130},
  year={2015}
}

@inproceedings{spray_and_wait_original,
author = {Spyropoulos, Thrasyvoulos and Psounis, Konstantinos and Raghavendra, Cauligi S.},
title = {Spray and wait: an efficient routing scheme for intermittently connected mobile networks},
year = {2005},
isbn = {1595930264},
publisher = {Association for Computing Machinery},
address = {New York, NY, USA},
url = {https://doi.org/10.1145/1080139.1080143},
doi = {10.1145/1080139.1080143},
booktitle = {Proceedings of the 2005 ACM SIGCOMM Workshop on Delay-Tolerant Networking},
pages = {252–259},
numpages = {8},
location = {Philadelphia, Pennsylvania, USA},
series = {WDTN '05}
}

@inproceedings{kim2010composite,
  title={Composite methods for improving spray and wait routing protocol in delay tolerant networks},
  author={Kim, Yong-Pyo and Koo, Ja-Il and Jung, Euihyun and Nakano, Keisuke and Sengoku, Mazakasu and Park, Yong-Jin},
  booktitle={2010 10th International Symposium on Communications and Information Technologies},
  pages={1229--1234},
  year={2010},
  organization={IEEE}
}

@misc{github_repo,
   author = {Ari Keränen},
   title = {GitHub - akeranen/the-one: The Opportunistic Network Environment simulator},
   url = {https://github.com/akeranen/the-one}
}

@misc{javadocs,
   title = {ONE Simulator Javadocs},
   url = {https://www.netlab.tkk.fi/tutkimus/dtn/theone/javadoc_v141/}
}

@misc{OpenStreetMap,
   author = {{OpenStreetMap contributors}},
   title = {{Planet dump retrieved from https://planet.osm.org }},
   howpublished = "\url{ https://www.openstreetmap.org }",
   year = {2017},
}

@misc{BMKG2022,
   author = {BMKG},
   title = {Sistem Monitoring InaTEWS},
   url = {http://202.90.198.40/sismon-wrs/web/slmon},
   year = {2022}
}

@misc{GoogleMaps2025,
   author = {GoogleMaps},
   title = {Kantor BPBD PROVINSI SULTENG - Google Maps},
   url = {https://www.google.com/maps/place/Kantor+BPBD+PROVINSI+SULTENG/@-0.89156,119.876373,3583m/data=!3m1!1e3!4m6!3m5!1s0x2d8bedacd6762c81:0x11c3c33f16a2b9e4!8m2!3d-0.8915598!4d119.8763725!16s%2Fg%2F11j54q_gny?hl=id&entry=ttu&g_ep=EgoyMDI1MTEyMy4xIKXMDSoASAFQAw%3D%3D},
   year = {2025}
}

@misc{BPBD2025,
   author = {BPBD},
   title = {BPBD PROVINSI SULAWESI TENGAH},
   url = {https://bpbd.sultengprov.go.id/},
   year = {2025}
}

@misc{IndonesiaStatePolice2022,
   author = {Indonesia State Police},
   title = {Number of Motor Vehicle by Type - Statistical Data - BPS-Statistics Indonesia},
   url = {https://www.bps.go.id/en/statistics-table/2/NTcjMg==/number-of-motor-vehicle-by-type--unit-.html},
   year = {2022}
}

@misc{InterimPopulationProjection2022,
   author = {Interim Population Projection},
   title = {Mid Year Population - Statistical Data - BPS-Statistics Indonesia},
   url = {https://www.bps.go.id/en/statistics-table/2/MTk3NSMy/mid-year-population--thousand-people-.html},
   year = {2022}
}

@misc{PaluCityCentralStatisticsAgency2023,
   author = {Palu City Central Statistics Agency},
   isbn = {2502-2881},
   month = {11},
   title = {Palu Municipality Regional Statistics 2022/2023 - BPS-Statistics Indonesia Palu Municipality},
   url = {https://palukota.bps.go.id/en/publication/2023/11/01/9310b1820947efe6e44fd3e3/palu-municipality-regional-statistics-2022-2023.html},
   year = {2023}
}

@article{al2023bluetooth,
  title={Bluetooth low energy for internet of things: review, challenges, and open issues},
  author={Al-Shareeda, Mahmood A and Saare, Murtaja Ali and Manickam, Selvakumar and Karuppayah, Shankar},
  journal={Indonesian Journal of Electrical Engineering and Computer Science},
  volume={31},
  number={2},
  pages={1182--1189},
  year={2023}
}

@article{koulouras2025evolution,
  title={Evolution of Bluetooth Technology: BLE in the IoT Ecosystem},
  author={Koulouras, Grigorios and Katsoulis, Stylianos and Zantalis, Fotios},
  journal={Sensors (Basel, Switzerland)},
  volume={25},
  number={4},
  pages={996},
  year={2025}
}

@inproceedings{yuliatmoko2023seismic,
  title={Seismic Station Quality Monitoring and Evaluation System in Indonesia},
  author={Yuliatmoko, RS and Gunawan, MT and Adi, M and Wijaya, A and Kambali, RAP and Kurniawan, T and Karnawati, DK and Rohadi, S},
  booktitle={IOP Conference Series: Earth and Environmental Science},
  volume={1276},
  number={1},
  pages={012045},
  year={2023},
  organization={IOP Publishing}
}

@misc{KennethAlambra2024,
   author = {Kenneth Alambra},
   month = {7},
   title = {Image File Size Calculator},
   url = {https://www.omnicalculator.com/other/image-file-size},
   year = {2024}
}

@book{pngs,
author = {Roelofs, Greg and Koman, Richard},
title = {PNG: The Definitive Guide},
year = {1999},
isbn = {1565925424},
publisher = {O'Reilly \& Associates, Inc.},
address = {USA}
}

@article{yang2023compression,
  title={Compression Performance Analysis of Different File Formats},
  author={Yang, Han and Qin, Guangjun and Hu, Yongqing},
  journal={arXiv preprint arXiv:2308.12275},
  year={2023}
}

@inproceedings{car_evacuation_feasibility,
author = {Sugiura, Shun and Fukai, Moeko and Miki, Saeko and Minegishi, Yoshikazu and Moriyama, Shuji and Hasemi, Yuji},
year = {2016},
month = {10},
pages = {pp.113-118},
title = {Feasibility of Car Evacuation from Tsunami in the Coastal Region},
journal = {Journal of Asian Urban Environment, Special Issue AIUE2016 Green Building and Smart City}
}

@inproceedings{maxprop,
author = {Burgess, John and Gallagher, Brian and Jensen, David and Levine, Brian},
year = {2006},
month = {04},
pages = {},
title = {MaxProp: Routing for Vehicle-Based Disruption-Tolerant Networks},
volume = {6},
journal = {Proceedings of IEEE INFOCOM},
doi = {10.1109/INFOCOM.2006.228}
}

@InProceedings{prophet,
author="Lindgren, Anders
and Doria, Avri
and Schel{\'e}n, Olov",
editor="Dini, Petre
and Lorenz, Pascal
and de Souza, Jos{\'e} Neuman",
title="Probabilistic Routing in Intermittently Connected Networks",
booktitle="Service Assurance with Partial and Intermittent Resources",
year="2004",
publisher="Springer Berlin Heidelberg",
address="Berlin, Heidelberg",
pages="239--254",
isbn="978-3-540-27767-5"
}

@article{MAKINOSHIMA2020100113,
title = {Tsunami evacuation processes based on human behaviour in past earthquakes and tsunamis: A literature review},
journal = {Progress in Disaster Science},
volume = {7},
pages = {100113},
year = {2020},
issn = {2590-0617},
doi = {https://doi.org/10.1016/j.pdisas.2020.100113},
url = {https://www.sciencedirect.com/science/article/pii/S2590061720300508},
author = {Fumiyasu Makinoshima and Fumihiko Imamura and Yusuke Oishi}
}

@INPROCEEDINGS{dtn_for_evacuation,
  author={Naito, Hikaru and Takai, Mineo and Ishihara, Susumu},
  booktitle={2024 International Conference on Information Networking (ICOIN)}, 
  title={Effect of Aggressive Evacuation Behavior Associated with Information-Sharing Using a DTN on Evacuation Time}, 
  year={2024},
  volume={},
  number={},
  pages={434-439},
  doi={10.1109/ICOIN59985.2024.10572130}
}

@INPROCEEDINGS{uav_dtns,
  author={de Albuquerque, José Carlos and de Lucena, Sidney C. and Campos, Carlos A. V.},
  booktitle={2016 IEEE 19th International Conference on Intelligent Transportation Systems (ITSC)}, 
  title={Evaluating data communications in natural disaster scenarios using opportunistic networks with Unmanned Aerial Vehicles}, 
  year={2016},
  volume={},
  number={},
  pages={1452-1457},
  doi={10.1109/ITSC.2016.7795748}
}

@Article{Bhattacharjee2015,
author={Bhattacharjee, Suman
and Roy, Siuli
and Bandyopadhyay, Somprakash},
title={Exploring an energy-efficient DTN framework supporting disaster management services in post disaster relief operation},
journal={Wireless Networks},
year={2015},
month={4},
day={01},
volume={21},
number={3},
pages={1033-1046},
issn={1572-8196},
doi={10.1007/s11276-014-0836-5},
url={https://doi.org/10.1007/s11276-014-0836-5}
}

@misc{iconixar2025,
   author = {iconixar and Freepik and EhtishamAbid},
   title = {Flaticon Vector Icons and Stickers},
   url = {https://www.flaticon.com/},
   year = {2025}
}

@article{palu_before_after,
author = {Titov, Vasily},
year = {2021},
month = {04},
pages = {},
title = {Hard Lessons of the 2018 Indonesian Tsunamis},
volume = {178},
journal = {Pure and Applied Geophysics},
doi = {10.1007/s00024-021-02731-0}
}

@misc{Ontheworldmapcom2019,
   author = {Ontheworldmap.com},
   title = {Indonesia Map | Detailed Maps of Republic of Indonesia},
   url = {https://ontheworldmap.com/indonesia/},
   year = {2019}
}

@misc{Freeworldmaps,
   author = {FreeWorldMaps},
   title = {Sulawesi map (fomerly Celebes)},
   url = {https://www.freeworldmaps.net/asia/indonesia/sulawesi.html}
}

@misc{kawano2025delaytolerantnetworkingtsunamievacuation,
      title={Delay-Tolerant Networking for Tsunami Evacuation on the Small Island of Hachijojima: A Study of Epidemic and Prophet Routing}, 
      author={Keiya Kawano and Milena Radenkovic},
      year={2025},
      eprint={2601.00109},
      archivePrefix={arXiv},
      primaryClass={cs.NI},
      url={https://arxiv.org/abs/2601.00109}, 
}

@ARTICLE{towards_low_cost,
  author={Radenkovic, Milena and Crowcroft, Jon and Rehmani, Mubashir Husain},
  journal={IEEE Access}, 
  title={Towards Low Cost Prototyping of Mobile Opportunistic Disconnection Tolerant Networks and Systems}, 
  year={2016},
  volume={4},
  number={},
  pages={5309-5321},
  doi={10.1109/ACCESS.2016.2606501}
}

@INPROCEEDINGS{enabling_real_time_communications,
  author={Radenkovic, Milena and Ha Huynh, Vu San and John, Robert and Manzoni, Pietro},
  booktitle={2019 International Conference on Wireless and Mobile Computing, Networking and Communications (WiMob)}, 
  title={Enabling Real-time Communications and Services in Heterogeneous Networks of Drones and Vehicles}, 
  year={2019},
  volume={},
  number={},
  pages={1-6},
  doi={10.1109/WiMOB.2019.8923246}
}

@ARTICLE{adaptive_real_time_predictive,
  author={Radenkovic, Milena and Huynh, Vu San Ha and Manzoni, Pietro},
  journal={IEEE Access}, 
  title={Adaptive Real-Time Predictive Collaborative Content Discovery and Retrieval in Mobile Disconnection Prone Networks}, 
  year={2018},
  volume={6},
  number={},
  pages={32188-32206},
  doi={10.1109/ACCESS.2018.2840040}
}

@INPROCEEDINGS{novel_framework_large_scale_emergency_communications,
  author={Ha Huynh, Vu San and Radenkovic, Milena},
  booktitle={2017 13th International Wireless Communications and Mobile Computing Conference (IWCMC)}, 
  title={A novel cross-layer framework for large scale emergency communications}, 
  year={2017},
  volume={},
  number={},
  pages={2152-2157},
  doi={10.1109/IWCMC.2017.7986616}
  }

\end{document}